\documentclass[9pt,twocolumn,twoside]{pnas-new}
\usepackage{units}
\usepackage{bm}
\usepackage{soul} 
\usepackage{verbatim}
\usepackage{xcolor,afterpage}
\usepackage{subfigure}
\usepackage{xspace}
\usepackage{aas_macros}
\usepackage[normalem]{ulem}
\newcommand{\HI}{H\textsc{i}\,\,}
\newcommand{\HII}{H\textsc{i}} 

\definecolor{darkgreen}{rgb}{0,0.6,0}

\newcommand{\beq}{\begin{equation}}
\newcommand{\eeq}{\end{equation}}
\def\Mpc{\, h^{-1} \, {\rm Mpc}}
\def\kMpc{\, h \, {\rm Mpc}^{-1}}

\def\Ms{\, h^{-1} \, M_\odot}
\def\mHI{M_\textup{HI}}
\defcitealias{VilGenCas1810}{VN18}

\templatetype{pnasresearcharticle} 
\setboolean{displaywatermark}{false}
\defcitealias{WadVil20}{W20} 

\title{Modeling assembly bias with machine learning and symbolic regression}

\author[a,1]{Digvijay Wadekar}
\author[b,c]{Francisco Villaescusa-Navarro}
\author[b,c,d]{Shirley Ho}
\author[c,e,f]{Laurence Perreault-Levasseur}

\affil[a]{Center for Cosmology and Particle Physics, Department of Physics, New York University, New York, NY 10003}
\affil[b]{Department of Astrophysical Sciences, Princeton University, Peyton Hall, Princeton NJ 08544-0010}
\affil[c]{Center for Computational Astrophysics, Flatiron Institute, 162 5th Avenue, 10010, New York, NY}
\affil[d]{Department of Physics, Carnegie Mellon University, Pittsburgh, PA 15217}
\affil[e]{Department of Physics, Universit\'e de Montr\'eal, Montr\'eal, Canada}
\affil[f]{Mila - Quebec Artificial Intelligence Institute, Montr\'eal, Canada}
\leadauthor{Wadekar} 

\correspondingauthor{\textsuperscript{1}Corresponding author. E-mail: jay.wadekar@nyu.edu}


\begin{abstract}

Upcoming 21cm surveys will map the spatial distribution of cosmic neutral hydrogen (H\textsc{i}) over unprecedented volumes. Mock catalogues are needed to fully exploit the potential of these surveys. Standard techniques employed to create these mock catalogs, like Halo Occupation Distribution (HOD), rely on assumptions such as the baryonic properties of dark matter halos only depend on their masses. In this work, we use the state-of-the-art magneto-hydrodynamic simulation IllustrisTNG to show that the H\textsc{i}\, content of halos exhibits a strong dependence on their local environment. We then use machine learning techniques to show that this effect can be 1) modeled by these algorithms and 2) parametrized in the form of novel analytic equations. We provide physical explanations for this environmental effect and show that ignoring it leads to underprediction of the real-space 21-cm power spectrum at $k\gtrsim 0.05 \, h \, {\rm Mpc}^{-1}$ by $\gtrsim$10\%, which is larger than the expected precision from upcoming surveys on such large scales. Our methodology of combining numerical simulations with machine learning techniques is general, and opens a new direction at modeling and parametrizing the complex physics of assembly bias needed to generate accurate mocks for galaxy and line intensity mapping surveys.

\end{abstract}

\doi{\url{www.pnas.org/cgi/doi/10.1073/pnas.XXXXXXXXXX}}

\graphicspath{{./}{figures/}}

\begin{document}

\maketitle
\thispagestyle{firststyle}
\ifthenelse{\boolean{shortarticle}}{\ifthenelse{\boolean{singlecolumn}}{\abscontentformatted}{\abscontent}}{}

\dropcap{I}n the coming decade,
numerous astronomical surveys will gather vast amounts of data about the Universe. 
To understand all that can be learned about the contents and evolution of the Universe from this data, scientists make predictions using different astrophysical and cosmological parameters. They organize these forecasts into catalogs of mock data, which can later be compared to the astronomical survey observations to infer the actual parameters describing the Universe. Astrophysicists use computer simulations of various types and levels of physical detail to build these catalogs.
In particular, there has been a lot of recent progress in developing
hydrodynamic simulations, which include the effects of 
star formation, gas cooling, magnetic fields, and energetic feedback due to supernovae and supermassive black holes.
That is, these simulations include detailed descriptions of both dark matter (DM), which has played a vital role in shaping the Universe's large-scale structure, and baryonic matter, which makes up the visible material in galaxies and galaxy clusters caught in the web of that structure.
The upcoming surveys, however, will span volumes of 10--100~(Gpc/h)$^3$, and the hydrodynamic 
simulations typically cost on the order of 10 million CPU hours for a mere $10^{-3}$~(Gpc/h)$^3$ \citep{PilSprNel1801}, 
making it impossible to use them directly for generating the required catalogs of large-scale mock data.
Less computationally expensive mock simulations will be needed to fill this gap.

One of the most popular theoretical techniques used to cheaply emulate the expensive hydrodynamic simulations and create large-scale mock baryonic data are halo models (also referred to as halo occupation distribution (HOD) models). HOD was first used to probabilistically model the number of galaxies residing in a host DM halo \citep{ScoSheHui01,Sel00,PeaSmi00,BerWei02} and it typically assumes a simple parametric relation between the halo's mass and its baryonic properties (e.g., the number of galaxies, stellar mass, or neutral hydrogen content of the halo). To calibrate the parameters in this relation, one typically uses semi-analytic models or hydrodynamical simulations (and observations, if available). The best-fit parametric relation is then applied to halos generated by gravity-only $N$-body simulations to make mock baryonic simulations.

HOD has frequently been used to make large-volume mock simulations, both for galaxy surveys \cite{ZheBer05, ReiSeo14, RodChu16, SaiLea16, AlaMiy17,HeaZen16,Avi20,DeRWec19,Zha19,LanVan19,YuaHadBos20,YuaEis20,WibWei20,AlaZu19,SalZu20} and for intensity mapping surveys \cite{VilVieDat14,VilVieAlo15,EmaPaco_17,VilGenCas1810,ModCasFen1909,Spi20}. Such mocks are used for 1) determining which summary statistics are the most appropriate to constrain different cosmological parameters; 2) studying the effect of various observational systematics on summary statistics and testing the range up to which perturbative models are robust for parameter inference; 3) constructing covariance matrices; and 4) simulation-based inference (SBI) analyses for parameter inference. 

The standard HOD technique assumes that the properties of various baryonic structures inside a halo are governed \emph{solely by the halo mass}, and ignores all other (``secondary'') halo properties. Other techniques used to make mocks rely on similar assumptions. For example, sub-halo abundance matching (SHAM; see e.g.~\cite{ValOst04,ConWec06}) assumes the existence of a scatter-free monotonic relation between
a halo's mass and the numbers or masses of the baryonic tracers in it.
However, numerous studies using hydrodynamical simulations and semi-analytic models have found that
the clustering of galaxies is in fact affected by secondary properties other than halo mass, such as the halo environment, halo assembly history, concentration, spin, velocity anisotropy, and many others \cite{Zhu06,PujGaz14,SchFre15,CroGaoWhi07,VakHah19, KobNisTak20, WecTin18,HadBosEis20}. This phenomenon is referred to as \emph{galaxy assembly bias}\footnote{
Note that galaxy assembly bias is different from halo assembly bias, which refers to the dependence of clustering of the DM halos themselves on secondary properties other than their mass \cite{Wec06, Dal08,ParHahShe18,HanLi19}. Halo assembly bias is automatically accounted for in an HOD analysis when halos from an $N$-body simulation are used. Note also that the term ``assembly bias'' was first used to characterize the effect of a particular secondary property---the halo assembly history \cite{SheTor04,Gao05}---but the term is now used more generally for the effect of all secondary properties of halos.}.

Studying galaxy assembly bias has been an important task as inaccurate galaxy mocks can lead to biases in the inferred cosmological parameters and galaxy formation properties.
There has also been some recent interest in this topic because of discrepancies in results inferred using the standard HOD model with the Planck cosmology on a simultaneous comparison to galaxy-galaxy lensing and projected galaxy clustering data from the BOSS CMASS and BOSS LOWZ samples \cite{LeaSai17, LanVan19,YuaEis20,YuaHadBos20,WibWei20,AmoBat20}.
Several studies have tried to incorporate secondary halo parameters into an HOD framework to include the effect of assembly bias. Halo concentration has been a traditionally popular secondary parameter \cite{CroGaoWhi07,VakHah19, KobNisTak20, WecTin18,ParKov15}, although numerous recent studies have shown that the environment of the halos plays a significant role for modeling the galaxy distribution \cite{McEWei18, XuZeh20, WibSalWei19, SalWibWei20,YuaHadBos20, HadBosEis20,HadBosEis20b}. 

All the works mentioned above have focused on understanding the galaxy-halo connection. However, with numerous upcoming surveys recording at the 21-cm wavelength (CHIME, HIRAX, HERA, TIANLAI, FAST, ASKAP, MeerKAT and SKA), which will probe the spatial distribution of neutral hydrogen (\HII) in the Universe, it now becomes imperative to also understand the connection between the properties of DM halos and their \HI content.

In this paper we quantify, for the first time, the effects of secondary properties of halos (i.e., properties other than their mass) on the clustering of \HI (i.e., \HI assembly bias) using the state-of-the-art IllustrisTNG magneto-hydrodynamic simulation. Naively, one might assume that baryons trace dark matter and therefore the \HI content of a halo should only depend on its total mass. However, we will show that this assumption is invalid and other halo properties such as the halo environment also have a crucial effect. Furthermore, also for the first time, we model \HI assembly bias using machine learning and symbolic regression in order to enable the creation of the more-accurate \HI catalogs needed to analyze data from the upcoming 21-cm surveys.

The paper is organized as follows. In sections~\ref{sec:Data} and~\ref{sec:HOD}, we describe the hydrodynamic simulation and the details of the HOD model that we use.
 In Sec.~\ref{sec:Env_mHI}, we quantify the effect of various halo secondary properties on halos' \HI masses and in its subsection~\ref{sec:physical} we discuss the physical reasons underlying these effects. In Sec.~\ref{sec:SyReg}, we model the \HII-halo connection using symbolic regression. In Sec.~\ref{sec:results}, we present the clustering of \HI from the different models. Finally, we discuss the relevance of our results, compare our approach to others in the literature, and conclude in Sec.~\ref{sec:Conclusions}.
 Before we begin our analysis, let us first discuss the motivation for using machine learning and symbolic regression to model the \HI assembly bias in the following three subsections.

\begin{figure}
\centering
\includegraphics[scale=0.4,keepaspectratio=true]{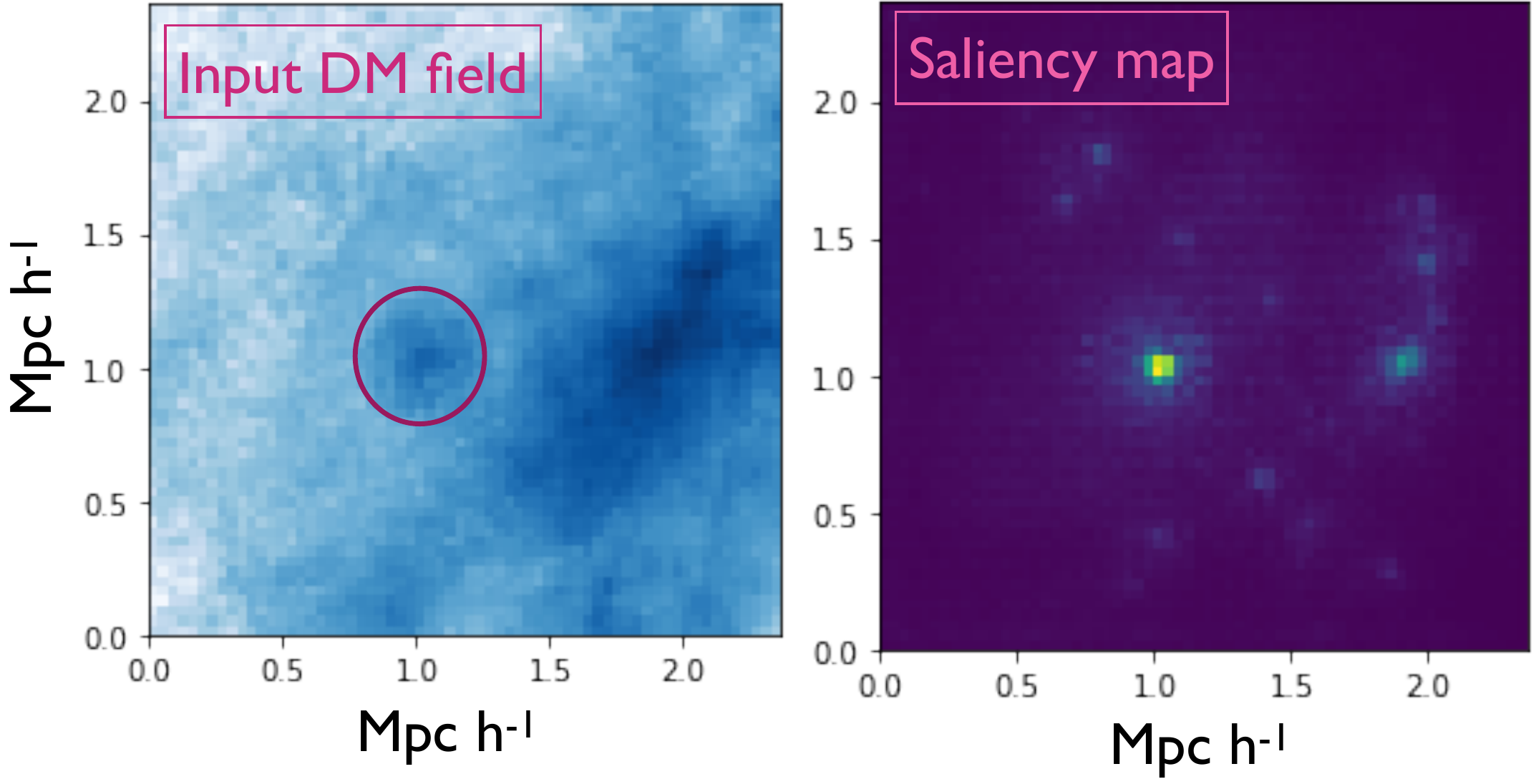}
\caption{
 \textbf{Left:} Dark matter (DM) density field in a particular sub-cube of the TNG100 simulation where we have identified a 10$^{12}\Ms$ halo (red circle). This field was provided as input to the U-Net trained in \citetalias{WadVil20} to predict the \HI field.
\textbf{Right:} Saliency map where brightness roughly corresponds to the importance of the input-field voxels used by the DNN to predict the \HI inside the circled halo. The bright regions well outside the halo indicate that the DNN used not only the local halo information, but also information in the halo's environment to make its prediction.
We find that the predicted \HI content in the circled halo is $\sim$15\% lower than if the same halo would be in an isolated environment.}
\label{fig:Saliency}
\end{figure}

\subsection{Motivation for halo environment from saliency maps of neural networks}
 Apart from theoretical techniques like HOD and SHAM which work on halo catalogs, machine learning tools like deep neural networks (DNNs) can be used to emulate expensive cosmological simulations directly at the field level \cite{WadVil20,ZhaWanZha1902,GuiReyVil1910,YipZha19,ZamOkaVil1904,HeLiFen1907,ModFenSel18,Kod20,TroFer19,Thiele_2020,LiNiCro20,BerSte19}.
A number of studies have shown that neural networks outperform traditional tools like HOD in emulating expensive simulations when various statistical properties of the emulated field are compared \cite{WadVil20,HeLiFen1907,GuiReyVil1910,ZhaWanZha1902,ZamOkaVil1904,YipZha19}. One question that arises here is whether there are any particular features in the input DM maps that the DNNs are using for their emulation and if such features can be used to augment standard HOD models.
However, one of the challenges to answer this question is that the DNNs are notoriously difficult to interpret;
this is due to the large number of fitted parameters (weights and biases) and also the depth of the many layers in a deep network.
There are however a few methods like saliency maps which can be used for getting insights into deep learning models \cite{ZorSha20,iNNvestigate} and we will discuss them below.

In our previous work, Ref.~\cite{WadVil20} (hereafter \citetalias{WadVil20}), we used a convolutional DNN to model the \HI field from an input matter field and showed that it outperforms HOD for emulating all summary statistics of the output \HI field ($\sim 15$\% improvement for the \HI power spectrum upto non-linear scales $k\leq1\kMpc$). As an the input to the DNN, \citetalias{WadVil20} used a high-resolution 3D matter field over a cube with side length $2.34 \Mpc$. For modeling \HI in a halo in the input field, the DNN therefore has access to information like the local environment of the halo and also the mass distribution inside the halo.
We are interested in roughly inferring what information is used by the DNN to make its prediction for \HII. To answer this question, we show, as an example, the saliency map corresponding to a particular case when a DM halo is in a tidal environment in Fig.~\ref{fig:Saliency}. 
One can visually see that the DNN models the information in the environment of the halo and uses it when predicting the \HI inside the circled halo. Furthermore, it is extremely interesting to see that the DNN lowers the \HI content of a halo when the halo is placed in an extremely overdense environment. This is analogous to the astrophysical effect called ram pressure stripping where gas escapes galaxies which are in a dense cluster due to the pressure from the surrounding ionized medium \cite{GunGot72}.

One question that still arises at this point is, what fraction of the network prediction comes by looking at the matter distribution outside halos against the one coming from inside the halo?
Answering this question using the DNN is a non-trivial task, given the complex nature of the flow of information within a typical DNN.
Furthermore, a common issue with DNNs is their generalizability, i.e their use on datasets with different hyperparameters than the ones they are trained on (for e.g. on an input field with a different resolution than the training sample). There is also a problem of data sparsity when using DNNs directly at the field level: most of the cosmological information comes from halos, which are found rarely in the input matter field \cite{WadVil20,ZhaWanZha1902,ModFenSel18,YipZha19}.


\subsection{Random forests for modeling assembly bias}
\label{sec:RF_intro}
In this paper we perform our analysis directly on the DM halo catalog rather than working on the field level. This drastically reduces the size of the dataset and therefore enables us to use traditional machine learning techniques like random forest regressors (RF), which are relatively less expensive and more interpretable than DNNs. One other advantage is that we can interface with the huge amount of theoretical work done
on halo models and we can quantify and understand the effect of various halo properties on its \HI mass ($M_\textup{HI}$).
Our goal is in this paper is to model $M_\textup{HI}$ by approximating the function
\beq
M_\textup{HI} = f(M_h, \{i_h\})
\label{eq:intro}\eeq
where $M_h$ is the total mass of the halo, and $\{i_h\}$ corresponds to the set of various secondary halo properties: the overdensity and anisotropy of its environment at various scales, concentration, assembly history like, formation epoch, spin, velocity dispersion...etc.
The high dimensionality of the input space makes this a complex and challenging problem. There are also correlations between different input parameters (e.g., the concentration of a halo is related to its assembly history and also its environment), which add to the complexity. 
 Machine learning tools like RFs are well-suited for approximating functions in a high-dimensional input parameter space; the advantage of using these methods over traditional theoretical methods is that there is no need to know the underlying functional form; only samples from that function are needed.

\subsection{Symbolic regression for parameterizing assembly bias}
\label{sec:SR_intro}
Symbolic regression (SR) is a technique that approximates the relation between an input and an output through analytic mathematical formulae \cite{SchLip09,UdrTeg20,WuTeg18,CraSan20,CraXu19, TorVilinprep,VilAngGen20,KimLu19,LiuTeg11}.
The advantage of using SR over other machine learning regression models 
is that it provides analytic expressions which can be readily generalized and which facilitate the  understanding of the underlying physics. One of the downsides of SR, however, is that the dimensionality of the input space needs to be relatively small.
In our case, we first use RF to get an indication of which parameters in the set of $\{i_h\}$ in Eq.~\ref{eq:intro} have the largest effect on $M_\textup{HI}$. We then compress the $\{i_h\}$ set to include only the five most important parameters. Finally, in Sec.~\ref{sec:SyReg}, we use SR on the compressed set to obtain an explicit functional form to approximate $f$ from Eq.~\ref{eq:intro}.
Having an analytic form with a minimal set of parameters that captures assembly bias is crucial in order to detect this effect through Bayesian analysis of real survey data. 
In this study, we use the symbolic regressor based on genetic programming implemented in the publicly available \textsc{PySR} package\footnote{\label{PySR}\href{https://github.com/MilesCranmer/PySR}{
\textcolor{blue}{https://github.com/MilesCranmer/PySR}}} \cite{pysr,CraSan20}. We leave further details on RF and SR to Appendix~\ref{apx:Techniques}.


\section{Data}
\label{sec:Data}
We use data from the TNG300-1 simulation produced by the IllustrisTNG collaboration \citep{PilSprNel1801}\footnote{\href{https://www.tng-project.org/data/}{\textcolor{blue}{https://www.tng-project.org/data/}}} throughout this paper. This simulation is a state-of-the-art magneto-hydrodynamic simulation that includes a wide range of relevant physical effects, such as radiative cooling, star formation, metal enrichment, supernova and AGN (active galactic nuclei) feedback, and magnetic fields. We will use two redshifts in our analysis: $z=5$  and $z=1$, corresponding to early and late times in the post-reionization era. It is worth mentioning that IllustrisTNG has already been used in multiple studies of galaxy assembly bias \citep{Mon20, ConAngZen20,ShiKurTak20,BosEis19,HadBosEis20,HadBosEis20b}. We show our results for the TNG300-1 simulation as it covers the largest volume among the TNG boxes. We also performed the analysis for the TNG100-1 simulation, which has a higher resolution than TNG300-1 but a smaller box size; we have checked that the \HI assembly bias results for TNG300-1 are qualitatively similar to the ones from TNG100-1, and therefore our results are robust against volume and resolution effects. 

\begin{figure}
\centering
\includegraphics[scale=0.33,keepaspectratio=true]{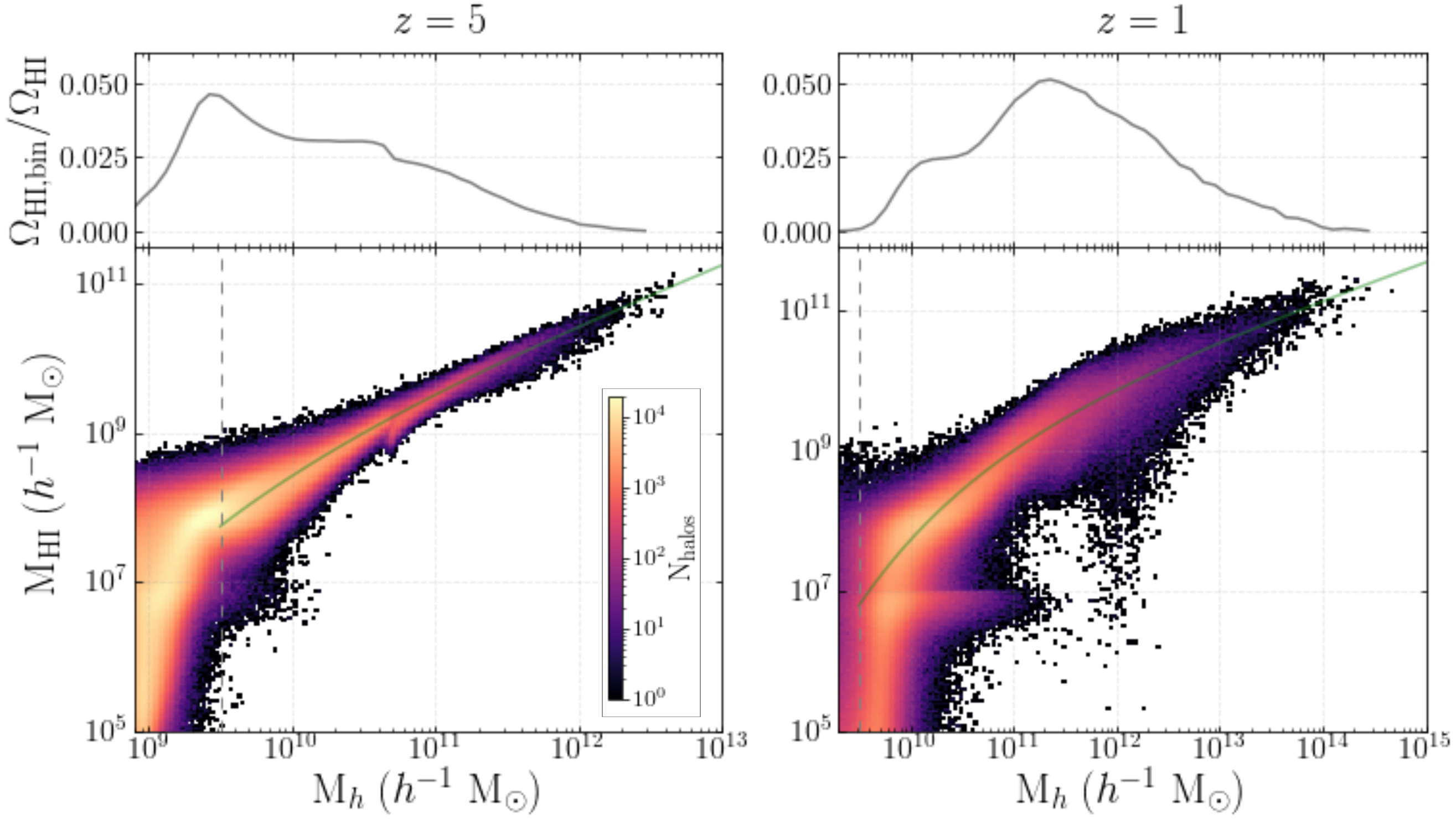}
\caption{\textbf{Bottom:} Distribution of \HI mass ($M_\textup{HI}$) versus total mass ($M_h$) for halos in the TNG300-1 simulation at redshifts $z=5$ (\emph{left}) and $z=1$ (\emph{right}). Color coding indicates the number of halos in that region of parameter space. The best-fit for the \HII--halo mass function (see Eq.~\ref{eq:HOD}) is shown in green and corresponds to the mass-only HOD prediction in the figures below. The dashed gray lines indicate the mass cutoff used in our analysis, that represents the mass of halos with $\sim$50 DM particles. \textbf{Top:} The fractional contribution to $\Omega_\textup{HI}$ of each halo mass bin.}
\label{fig:HImassfn}
\end{figure}

\section{HOD model for \HI}
\label{sec:HOD}

Halo models have been traditionally popular for modeling galaxies and there have been multiple recent attempts at developing a halo model for the abundance and spatial distribution of \HI \cite{VilVieDat14,VilVieAlo15,EmaPaco_17,VilGenCas1810, Spi20,PadChoRef16,PadRef17,PadRefAma17,PadRefAma19,PadRefAma20,BarHae14}.
The main idea behind such models is that most of the \HI in the post-reionization era resides inside halos: more than 99\% at $z<0.2$ (the fraction decreases to $\sim$88\% at $z=0.5$) \cite{VilGenCas1810}. We can use this fact to generate \HI fields by populating halos in an $N$-body simulation with \HI and this method is called as HOD.\footnote{HOD in the traditional literature is used for modeling the number of galaxies in a particular halo and we use the term here for modeling the mass of \HI in a particular halo.}.
In this study, we will use the HOD model of Ref. \cite{VilGenCas1810} (hereafter \citetalias{VilGenCas1810}), which has also been used to make gigaparsec volume \HI mocks \cite{ModCasFen1909}.
We briefly describe their model in what follows and refer the reader to \citetalias{VilGenCas1810} for further details.
The first step involves running a DM-only simulation, identifying halos, and saving their positions and masses.
A DM halo of mass $M_h$ is then assigned an \HI mass (denoted by $M_\textup{HOD-HI}$) using the \HII$-$halo mass relation given by:
\beq
M_\textup{HOD-HI} (M_h,z) = M_0 \left(\frac{M_h}{M_\textup{min}}\right)^\alpha \exp [-(M_\textup{min}/M_h)^{0.35}]
\label{eq:HOD}\eeq
where $M_0$ is a normalization factor, $\alpha$ is the power-law slope, and $M_\textup{min}$ is the characteristic minimum mass\footnote{It gets harder to retain neutral gas in halos below this mass which can self-shield itself from the ionizing metagalactic radiation.} of halos that host \HII. We calibrate this relation using halos from the TNG300-1 simulation; we only consider halos with masses above $10^{9.5} \Ms$, which have $\sim 50$ bound DM particles, to ensure our sample is robust.
We get the best-fit values for $z=1$ $(z=5)$ to be: $M_0 = 0.64\, (0.2) \times 10^{10} \Ms$, $M_\textup{min}=  25.41\, (2.36) \times 10^{10} \Ms$ and $\alpha=0.52\, (0.76)$. We show the corresponding fits in Fig.~\ref{fig:HImassfn}. Note that the best-fit parameter values are different from those in \citetalias{VilGenCas1810}, which were calibrated for the TNG100-1 simulation. This is caused by 1) resolution effects, that affect the strength of the astrophysical effects such as AGN and supernova feedback, and 2) the different choice of halo mass cutoff for calibrating the $M_\textup{HOD-HI}$ relation. One can immediately note from Fig.~\ref{fig:HImassfn} that there is a large scatter in the \HI-halo mass relation at fixed $M_h$. We will later show that a part of this scatter is due to halo environment and discuss its impact on the clustering of the modeled \HI field. 
As we are interested in analyzing large scales in the paper, we have ignored the one-halo term (which account for distribution of \HI within the halo) in our model throughout the paper and assumed the entire \HI is located at the center of halo; the one halo term only becomes important on scales $k\gtrsim 1 \kMpc$ that are not relevant for this work.


We emphasize that the best-fit values quoted earlier for $\{M_0,\alpha,M_\textup{min}\}$ were obtained by using halo parameters from the hydrodynamical (or full physics, FP) IllustrisTNG simulation and therefore these best-fit values \emph{should not be directly used on halos in a $N$-body (dark matter only, DMO) simulation}. This is because baryonic effects have a significant effect on the halo mass \citep{LovPilGen18,SchFre15}. Using the value of the free parameters calibrated from the FP simulation, directly for a $N$-body simulation, would lead to significant discrepancies: for e.g., $\Omega_\textup{HI}$ will change by $\sim$15\%. We have separately calibrated the fitting formula in Eq.~\ref{eq:HOD} using halos from the phase matched DMO version of the TNG300-1 simulation and we present the corresponding best-fit parameters in Appendix~\ref{apx:Halos_FPvsDMO}.
We leave further discussion of the differences in the halo masses between the FP and DMO simulations for Fig.~\ref{fig:DMOvsFP} and Appendix~\ref{apx:Halos_FPvsDMO}.


\section{Effect of secondary halo properties on its \HI mass}
\label{sec:Env_mHI}

\begin{figure*}
\centering
\includegraphics[scale=0.43,keepaspectratio=true]{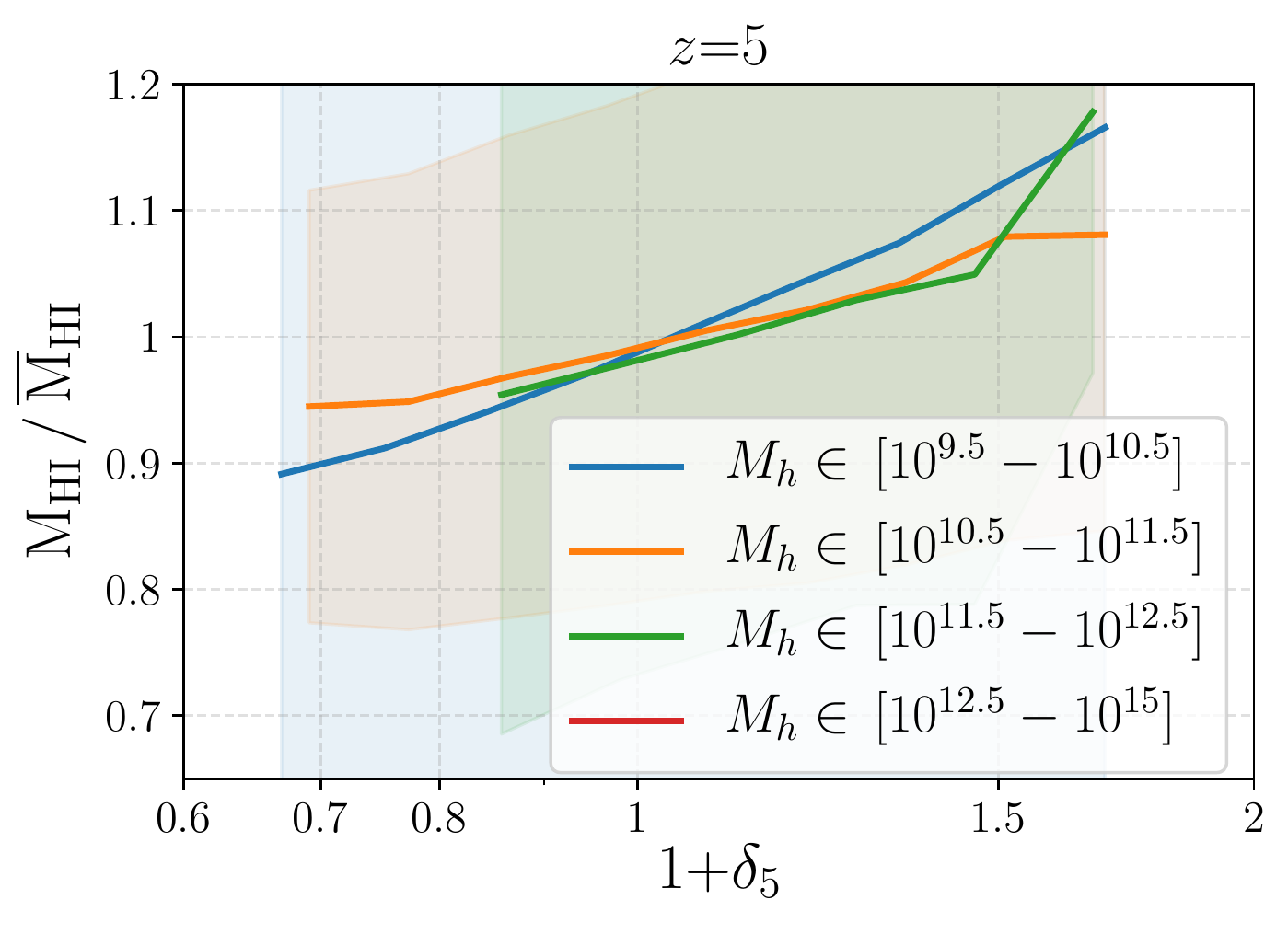}
\includegraphics[scale=0.43,keepaspectratio=true]{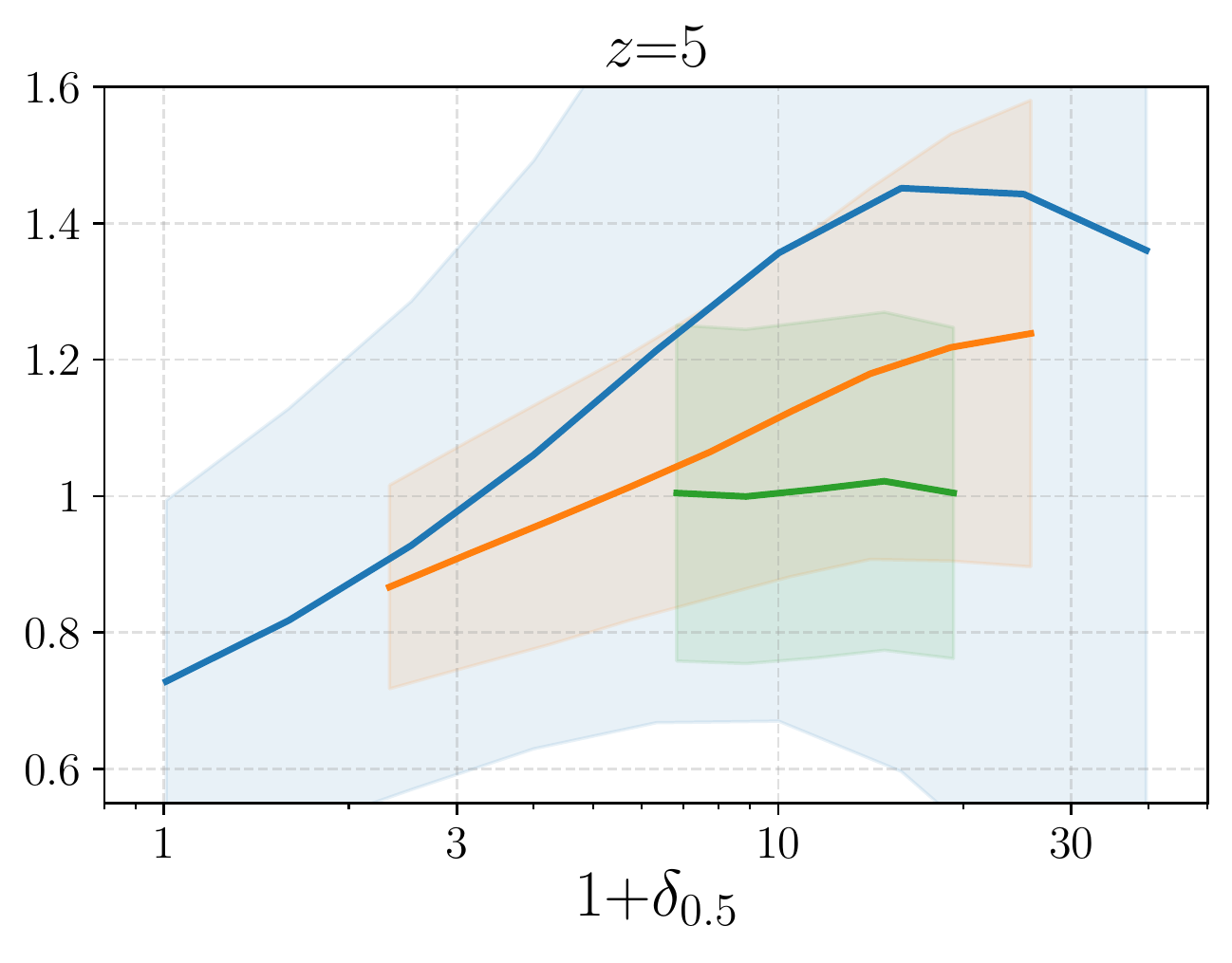}
\includegraphics[scale=0.43,keepaspectratio=true]{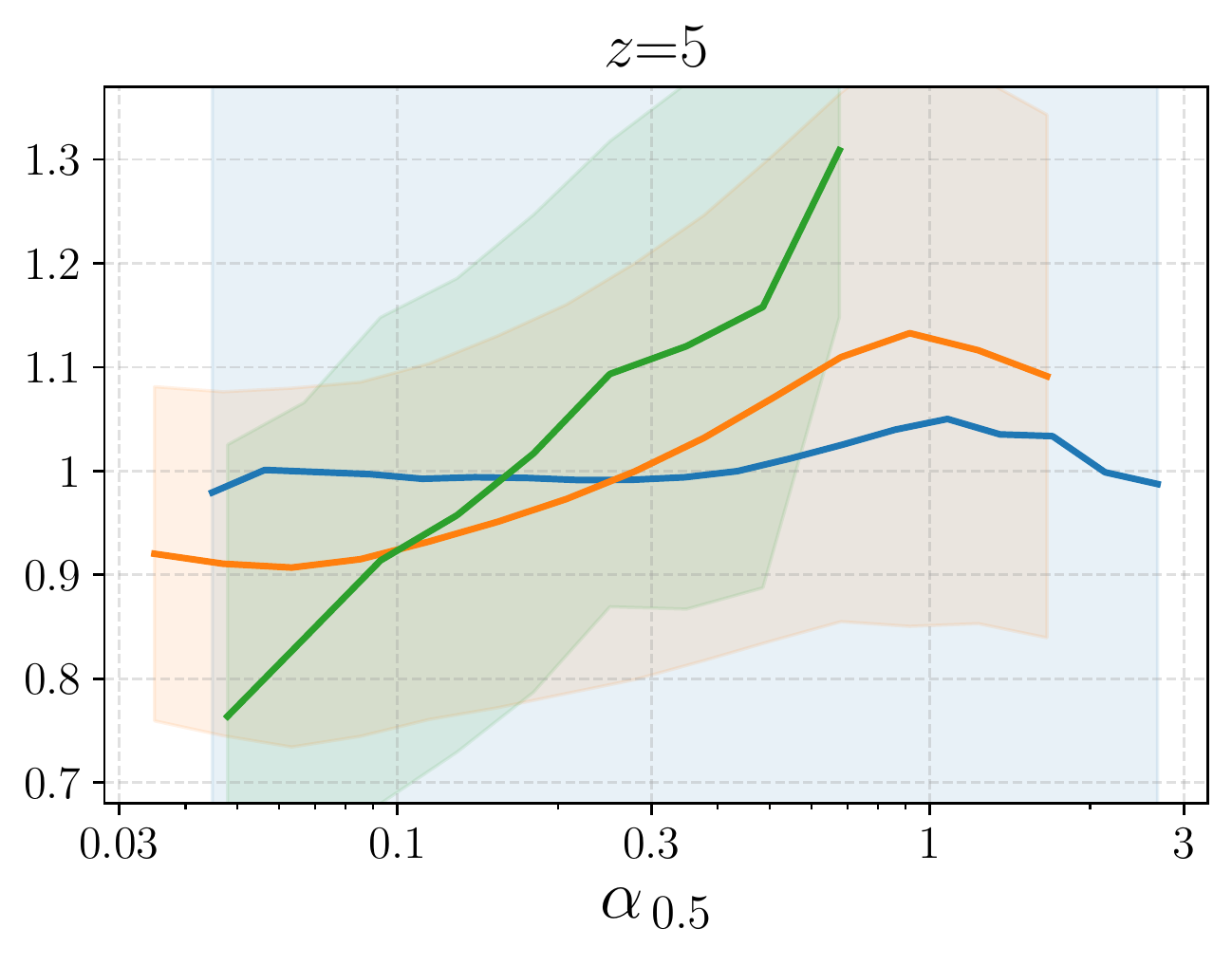}
\includegraphics[scale=0.43,keepaspectratio=true]{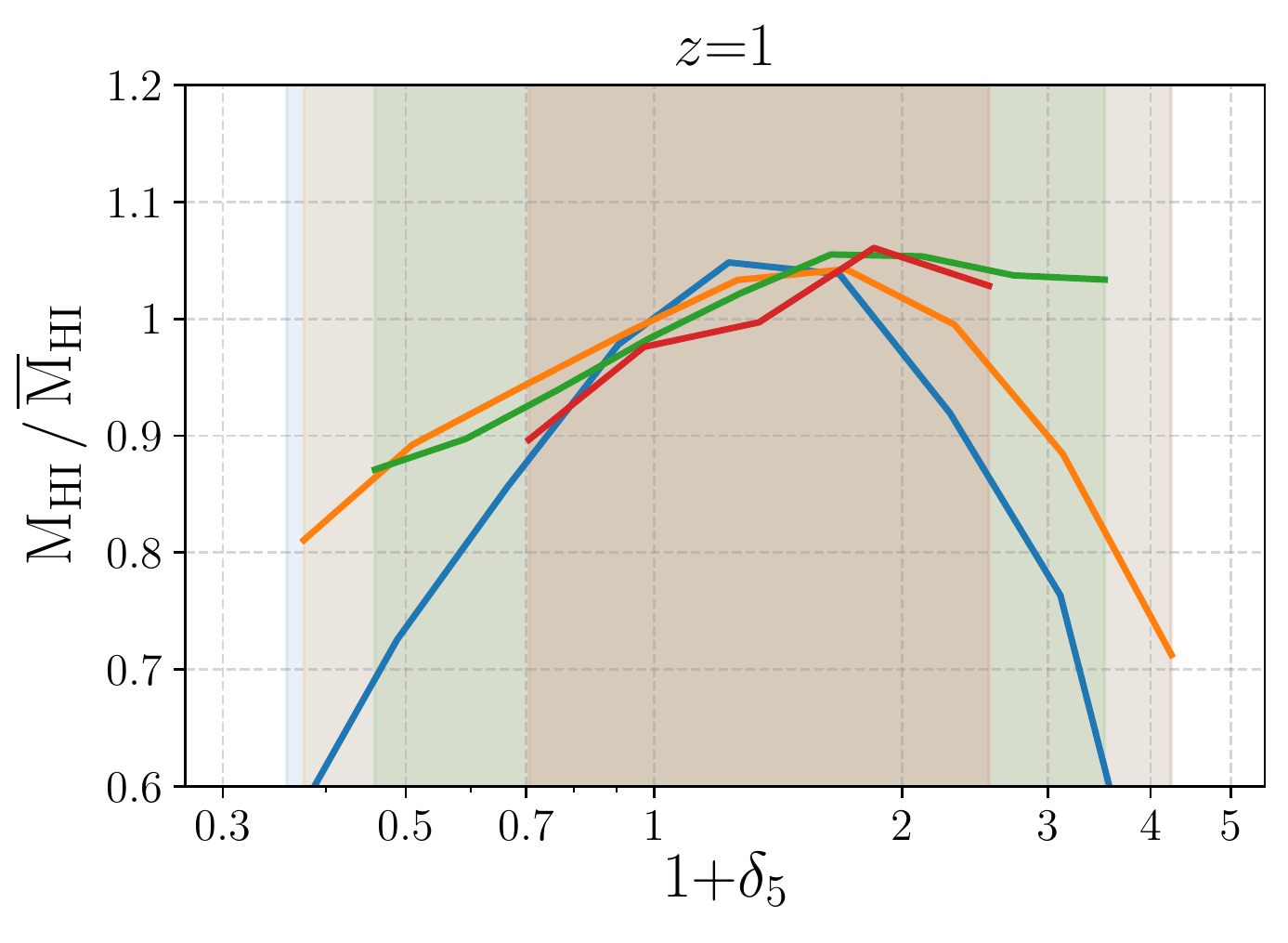}
\includegraphics[scale=0.43,keepaspectratio=true]{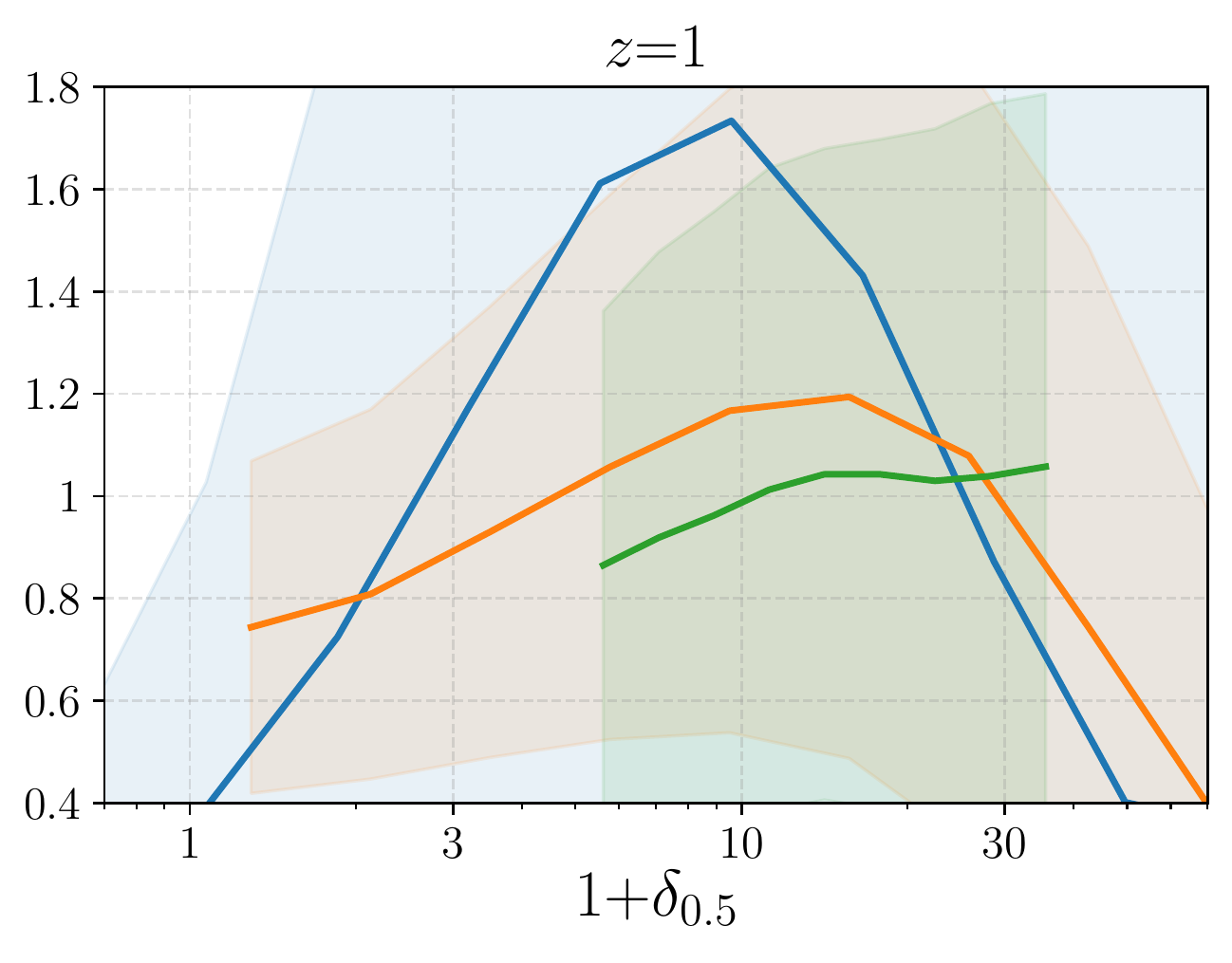}
\includegraphics[scale=0.43,keepaspectratio=true]{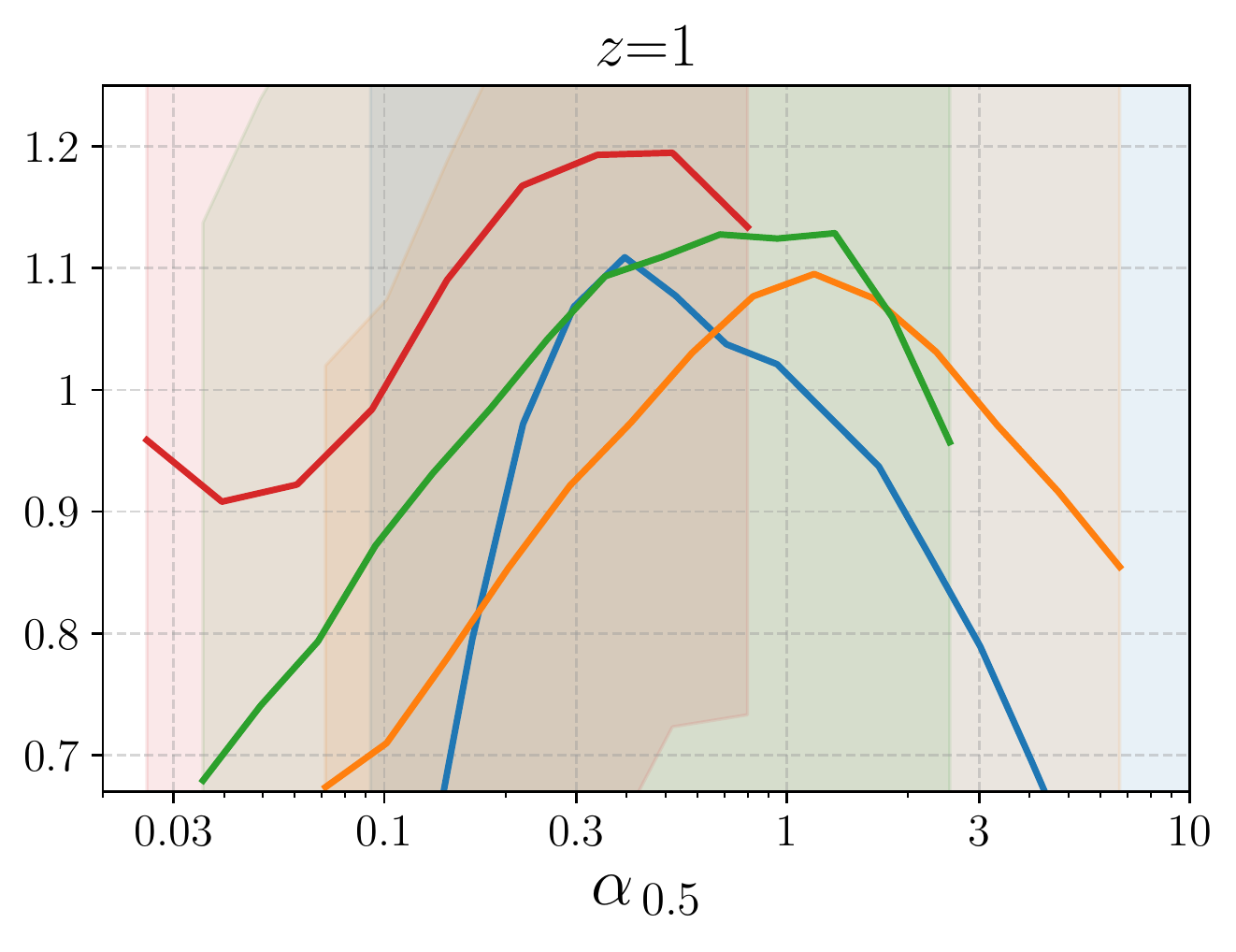}

\caption{This figure shows the effect of various halo environmental properties on their \HI masses ($\textup{M}_\textup{HI}$). $\delta_R$ ($\alpha_R$) is the environmental overdensity (anisotropy) within a radius $R/(\Mpc)$ surrounding the halo as defined in Eq.~\ref{eq:envDelta} (Eq.~\ref{eq:alpha}). We split the halos in the TNG300-1 volume into four mass bins and show the ratio of their \HI mass to the average \HI mass of halos in the corresponding mass-bin; solid lines are the mean relations and shaded areas enclose 68\% of the data. If $\textup{M}_\textup{HI}$ would only depend on $M_h$, the result will be constant value of one on the y-axis; we instead see strong mean trends for effect of the environment (especially for the low mass halos). 
There is also a large scatter, which is a result of the highly stochastic nature of gas accumulation in a halo.
See Sec.~3\ref{sec:physical} for a physical explanation of the mean trends. All masses are in units of $\Ms$.
}
\label{fig:mHI}
\end{figure*}
\begin{figure*}[h!]
\centering
\includegraphics[scale=0.43,keepaspectratio=true]{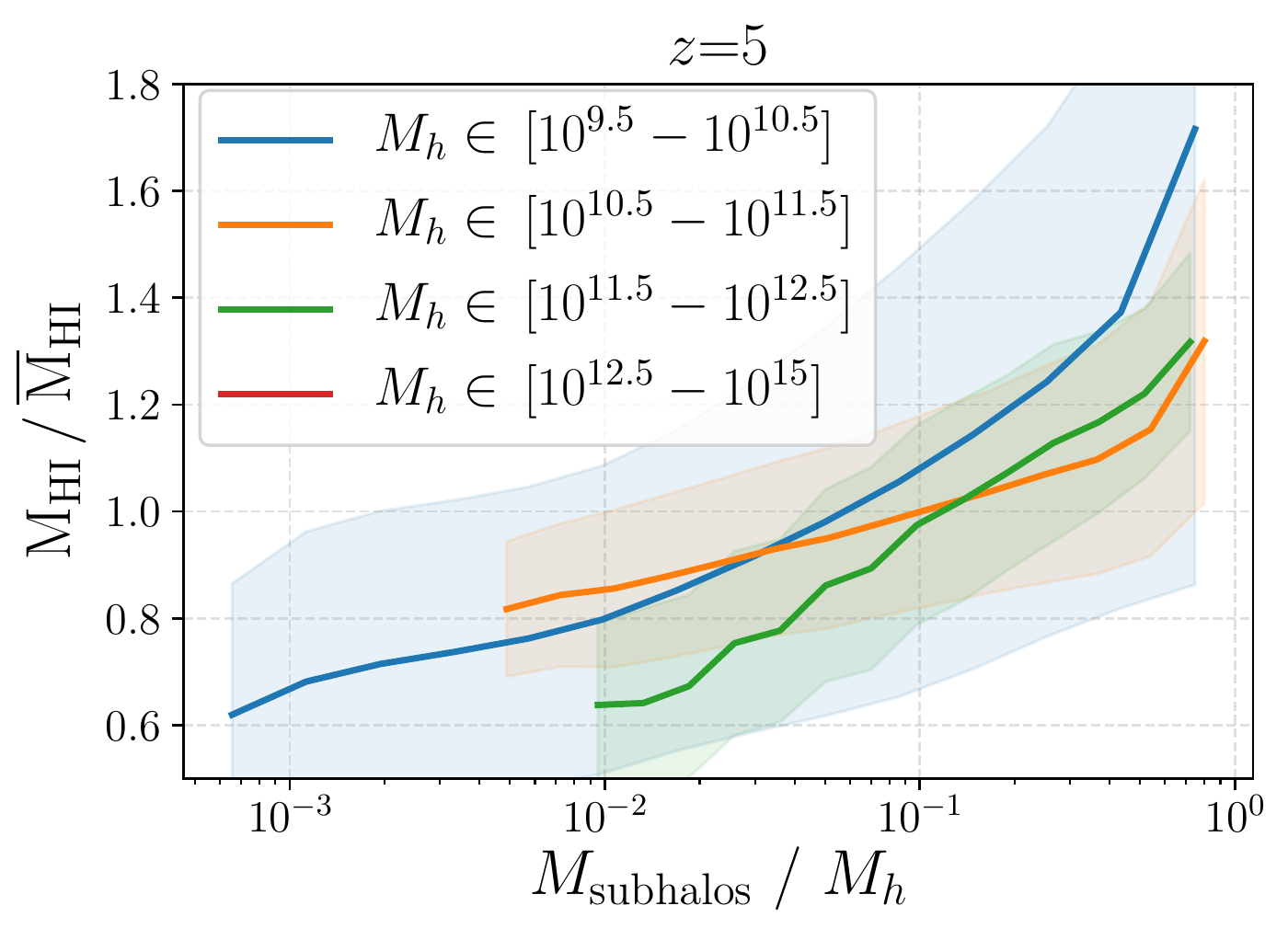}
\includegraphics[scale=0.43,keepaspectratio=true]{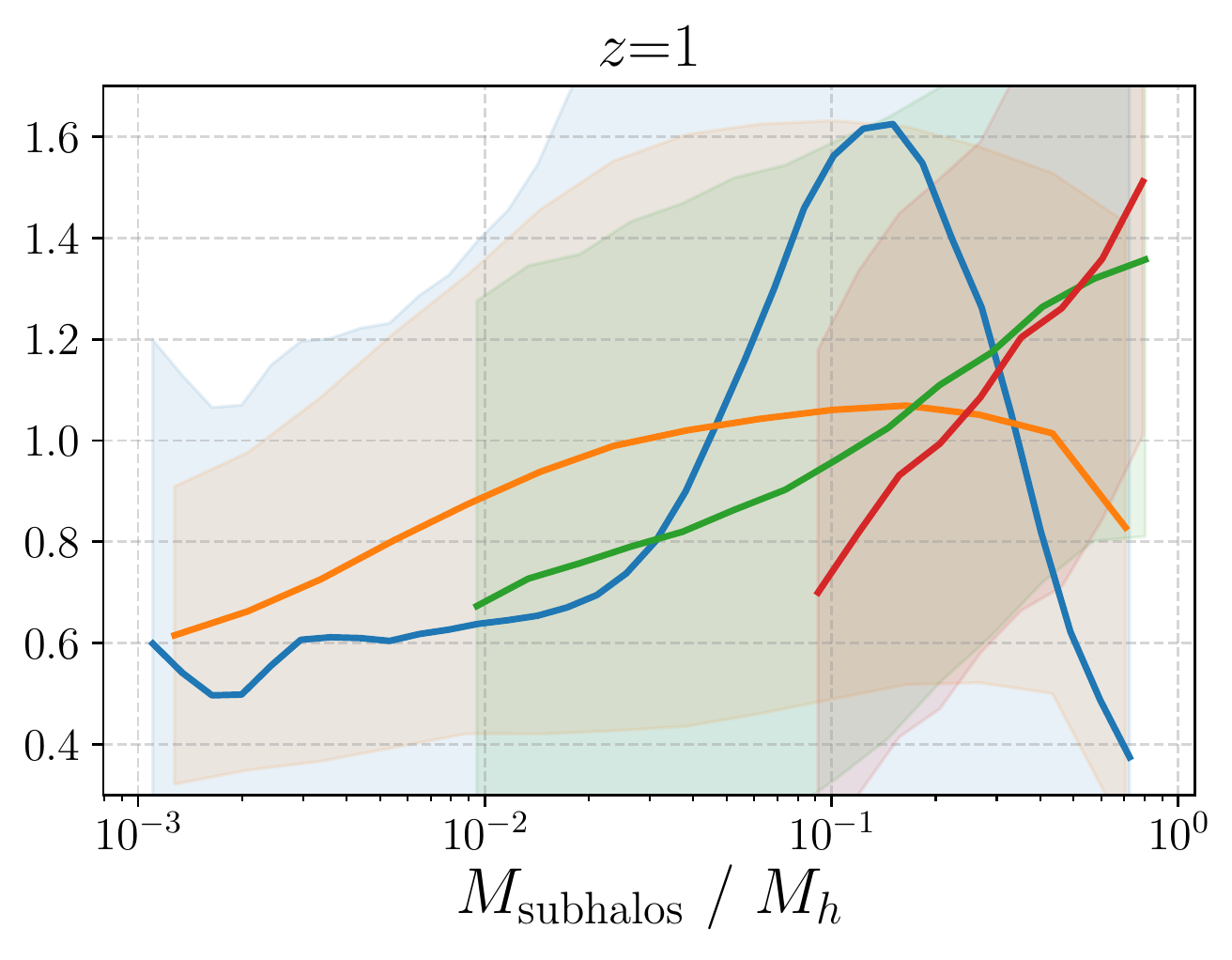}
\includegraphics[scale=0.43,keepaspectratio=true]{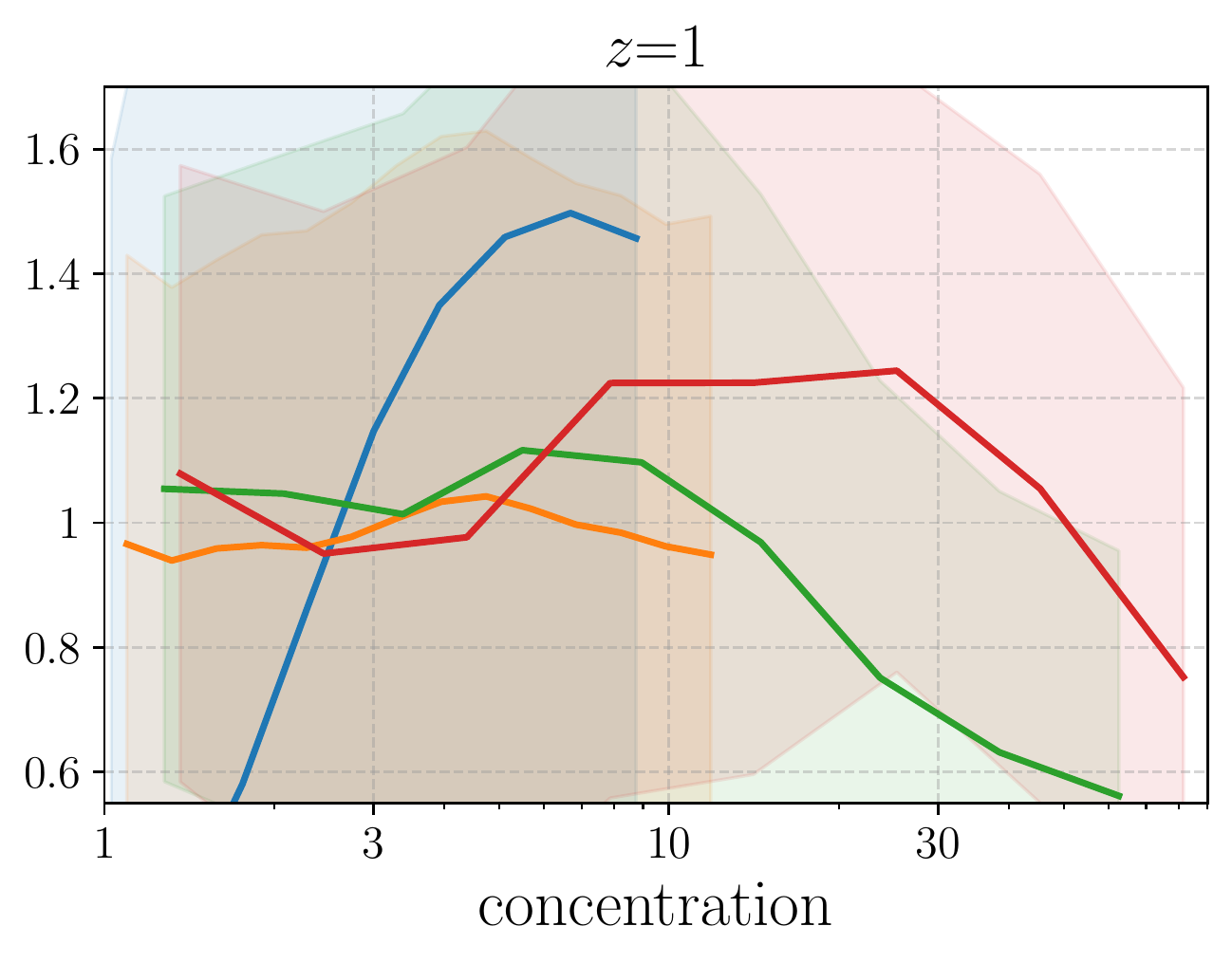}
\caption{Same as Fig.~\ref{fig:mHI} but for the effect of internal halo properties like the fractional mass in subhalos (left and middle) and concentration on the halo \HI mass (right). These trends are shown to complement the physical explanations outlined in Sec.~3\ref{sec:physical} for the effect of the environment seen on the halo \HI mass in Fig.~\ref{fig:mHI}.}
\label{fig:mHI2}
\end{figure*}

\citetalias{VilGenCas1810} showed that halos with the same mass but different \HI content cluster differently. This hints at the fact that the \HI content of the halo is affected by secondary properties of the halo rather than $M_h$ alone. We explore this effect in detail in this section.
We focus on 1) halo environmental parameters and 2) internal halo properties such as mass fraction in subhalos and concentration.

Let us first outline the procedure that we use to calculate the environmental overdensity ($\delta_R$) within a radius $R$ ($\Mpc$) surrounding the halos. 
We compute the matter overdensity field $\delta$ of the TNG300-1 box on a $N_g$ = $2048^3$ grid using CIC interpolation and then smooth in Fourier space by a top-hat filter with radius $R$.
We then transform the smoothed field back to real space and interpolate the smoothed field to the locations of the halos. We finally calculate the overdensity after subtracting the contribution from the halo mass, which is equivalent to using 
\beq
1+\delta_R \equiv \frac{1}{\bar{\rho}} \frac{M_R-M_h}{4/3\,\pi R^3}\, ,
\label{eq:envDelta}\eeq
where $M_R$ is the total mass within a radius $R$\footnote{We assign $\delta_R=0$ when $M_R<M_\textup{halo}$ which roughly happens when the $R<R_\textup{virial}$ for large halos.}.
 The subtraction is made to make $\delta_R$ independent of halo mass. 
It is worth mentioning that some studies in the literature use a Gaussian kernel instead of a top-hat kernel for smoothing the density field to calculate the effect of the environment. In that case, however, the environmental variable has some contribution from the halo mass itself and we choose our definition to make $\delta_R$ independent of halo mass.

Apart from the environmental overdensity, we also explore the dependence of $M_\textup{HI}$ on the anisotropy of the matter distribution around halos. There is increasing evidence that the tidal anisotropy is a key factor in determining halo assembly bias \cite{RamParHah19,ParHahShe18,ObuDalPer19,HahPor09,ManKra20}, and we therefore investigate whether it affects the \HI clustering.
In order to quantify the anisotropy, we first calculate a dimensionless version of the tidal tensor as $T_{ij}\equiv \partial^2 \phi_R/\partial x_i \partial x_j$, where $\phi$ is the dimensionless potential field calculated using Poisson's equation: $\nabla^2 \phi_R = -\rho_R/\bar{\rho}$.
We then calculate the tidal shear $q^2_R$ using\footnote{Note that perturbation theory based models use a closely related variable $s^2\equiv2q^2/3$ for studying the non-local bias \cite{ChaSco12,BalSel12}.} \cite{CatThe96, HeaPea98}
\beq
q^2_R\equiv \frac{1}{2} \big[ (\lambda_2-\lambda_1)^2+(\lambda_3-\lambda_1)^2+(\lambda_3-\lambda_2)^2\big]
\eeq
where $\lambda_i$ are the eigenvalues of $T_{ij}$. However, Ref.~\cite{ParHahShe18} showed that tidal shear on small scales is also correlated with the environmental overdensity and in order to isolate the anisotropy effect one needs to appropriately normalize the shear.
We adopt the normalization of the shear proposed by Ref.~\cite{ParHahShe18} which is 
\beq
\alpha_R \equiv \frac{\bar{\rho}}{\rho_R} \sqrt{q^2_R}
\label{eq:alpha}
\eeq
where the scalar parameter $\alpha_R$ is referred to as the tidal anisotropy parameter and efficiently encodes the tidal information. Ref.~\cite{ParHahShe18} also recommended to use an adaptive top-hat smoothing scale $4\, R_{200}$ for individual halos;  we however use a global smoothing scale for all halos as it is computationally more straightforward to calculate using fast fourier transforms (FFTs). The smallest scale for smoothing that we consider is $0.5 \Mpc$, as it becomes increasingly expensive to smooth to smaller scales because a larger grid with a finer resolution is needed.
We show the relation of $M_\textup{HI}$ with the halo environmental overdensity and anisotropy in Fig.~\ref{fig:mHI}. The mean of the trends are shown in solid and we find a large scatter in the relation as seen from the shaded areas representing 1$\sigma$ deviations (we will later see in Fig.~\ref{fig:Pow_AllEnv} and Sec.~\ref{sec:results} that the mean trends have a significant effect on the clustering statistics of \HI and the scatter largely gets averaged out). We pick two scales to show the dependence of the results with the smoothing scale: $R=0.5 \Mpc$ for the small-scale environment, and $R=5 \Mpc$ for the large-scale environment.
We did not show the trends for smoothing scales larger than $5 \Mpc$ because the effect of the halo environment starts to become less significant.
We do not show lines corresponding to high-mass halos ($M_h\gtrsim 10^{12.5}\Ms$) for $z=5$ (as such halos are rare at high redshifts) and for $z=1$ (as such halos dominate the small-scale environment and $\delta_{0.5}\rightarrow 0$ as per Eq.~\ref{eq:envDelta}). We also have not shown the number of halos in a particular region of the parameter space in Fig.~\ref{fig:mHI} (there are typically much fewer halos towards the high end of the environmental parameter values).

Overall, we see that the effect of halo environment is more pronounced for low-mass halos $M_h\lesssim 10^{11.5} \Ms$. 
Note that such low-mass halos do not typically host galaxies that can be observed in galaxy surveys: they would be too faint. On the other hand, these halos, due to the large abundance, host a very large fraction of the total \HI mass,
as seen in the top panel of Fig.~\ref{fig:HImassfn}.


\subsection{Physical explanations for the environmental trends}
\label{sec:physical}
Let us now provide a physical interpretation to some of the trends of $M_\textup{HI}$ with environmental variables in Fig.~\ref{fig:mHI}.
A key physical effect for understanding the \HI content of a halo is that, in the presence of ionizing radiation, \HI can only form in places where the gas has sufficiently high density to self-shield itself. Let us, for clarity, discuss explanations of the environmental trends for the two redshift cases separately:

 \textit{High-$z$ case} ($z=5$): Increasing the environmental overdensity typically results in more mergers and therefore more substructure in halos (see Fig.~\ref{fig:mHI_others}). We show in the left panel of Fig.~\ref{fig:mHI2} that, at fixed halo mass, increasing the fraction of mass in subhalos increases the \HI content of the halo. This is likely because the gas in subhalos is more dense and can more efficiently self-shield itself as compared to gas in the CGM (circumgalactic medium) of the halo.

\textit{Low-$z$ case} ($z=1$):
The main difference here, as compared to the high-$z$ case, is that the ionizing metagalactic background and the feedback due to AGN and supernovae are both much stronger, and lead to ionization of \HI in low density regions. Let us first discuss the high-mass end: $M_h>10^{12} \Ms$. Due to strong AGN feedback in the central galaxy, a significant fraction of \HI is located in the subhalos of these halos (see Fig.~7 of \citetalias{VilGenCas1810}). This explains why we find a correlation between the \HI mass and the total mass in subhalos (see central panel of Fig. \ref{fig:mHI2}).
This explanation breaks down for low-mass halos, where we see a turnover in the trends (especially for smaller halos) when the environment becomes dense or anisotropic beyond a certain threshold. The turnover likely occurs in cases when a halo is close enough to large objects like galaxy clusters; those halos can lose their gas content due to ram pressure stripping. It is worth noting that this effect is similar to the one in Fig.~\ref{fig:Saliency}, where the neural network lowers the \HI inside a halo when its is located in a highly overdense and anisotropic environment. This hints at the fact that neural networks can learn to model complex astrophysical effects directly from the output data of hydrodynamic simulations.

There is another effect of halo concentration which comes into play for the lowest mass halos $10^{9.5}<M_h<10^{10.5} \Ms$ at $z=1$. For these halos, the ionizing feedback from the central galaxy is much lower and most of the \HI is concentrated in the central galaxy. Having a higher concentration therefore makes it easier to accumulate a large density of gas in the center of the halo (which is required for \HI to self-shield itself from the metagalactic radiation), and hence we see the steep rising trend in the right panel of Fig.~\ref{fig:mHI2}. 
The low-mass halos in more dense and anisotropic environments typically have higher concentration \cite{Wec06,ParHahShe18}, which could be the reason behind the initial steep rising trend with the environment for such halos in Fig.~\ref{fig:mHI}.
Among the internal properties of the halo, we have only considered the effect of the subhalo mass-fraction and concentration in this work. We leave further discussion on the halo internal properties to Appendix~\ref{apx:OtherParams}, where we also show additional plots corresponding to these parameters.

\section{Modeling the halo HI mass with symbolic regression}
\label{sec:SyReg}
Our goal in this section is to model the trends observed in Fig.~\ref{fig:mHI} with compact analytic expressions using symbolic regression (SR). As discussed earlier in Sec.~\ref{sec:SR_intro}, SR allows us to obtain a functional form that can capture the structure in a high-dimensional dataset, and is therefore an ideal tool for our purposes. 

Before we input the various environmental parameters into the symbolic regressor, we rescale the parameters by using logarithms to shorten the range over which these parameters vary: $m_{10} \equiv \log[M_h /(10^{10} \Ms)]$ for the halo mass, $\delta'_\textup{R}\equiv\log (2+\delta_\textup{R})$ for the environmental overdensity and $\alpha'_{R}\equiv \log(1+\alpha_{R})$ for the environmental anisotropy\footnote{Note that the particular constants are added before using the logarithm in order to prevent the rescaled parameters from diverging when $\delta \rightarrow -1$ or $\alpha\rightarrow0$.}.
We use the SR for modeling the ratio of $M_\textup{HI}$ to the output $M_\textup{HOD-HI}$ from the mass-only HOD model from Eq.~\ref{eq:HOD}. 
We train the SR separately at two redshifts and get
\begin{subequations}\label{eq:SyReg}
     \begin{align}
      \frac{M_\textup{HI}}{M_\textup{HOD-HI}}
 =&\, 0.95+\alpha'_{0.5}\, \delta'_{0.5}\, (\alpha'_{0.5} + \delta'_{0.5}) \label{eq:SyRegz5} \quad\ \ [z=5]\\
     \frac{M_\textup{HI}}{M_\textup{HOD-HI}}
 =&\, 0.81\, + 1.44\, \alpha'_{0.5}\, m_{10} \nonumber \\
& -0.57\, (\alpha'^2_{0.5}\, m^2_{10} + \alpha'_{0.5}\, \delta'_{5}) \quad\ \ \ [z=1] 
 \label{eq:SyRegz1} \end{align}
\end{subequations}
We have presented the most concise expressions that include the effect of  environment on the \HI mass over the full range of halo masses.
It is important to note that the expressions in Eq.~\ref{eq:SyReg} are not unique, i.e we have found expressions which fit the TNG data better than the ones in Eq.~\ref{eq:SyReg}, however their form is relatively much more complex.
Furthermore, due to the presence of a large scatter in the environmental relations as seen in Fig.~\ref{fig:mHI}, the risk of overfitting goes up as the equations get more complex. 
A question which arises at this point is whether the forms of Eq.~\ref{eq:SyReg} are robust when the astrophysical feedback and cosmology parameters in the hydrodynamic simulations are changed. We plan to answer this in a future study using the CAMELS simulations suite \cite{VilAngGen20}, which contains multiple hydrodynamic simulations run with different feedback parameters.

Note that the expressions in Eq.~\ref{eq:SyReg} do not contain all the environmental parameters seen in Fig.~\ref{fig:mHI} (for e.g., Eq.~\ref{eq:SyRegz5} does not involve the large-scale term $\delta_5$); this is likely because the environment at different scales is correlated and sometimes the information gained from different environmental parameters is degenerate. We also show in Fig.~\ref{fig:mHI_SyReg} the performance of these expressions.
Let us discuss the connections of these equations to some of the trends seen in Fig.~\ref{fig:mHI}. For $z=1$, apart from a constant, there is a linear term with respect to the environmental variables ($1.44\, \alpha'_{0.5}\, m_{10}$) and a corresponding quadratic order term with a negative sign. The negative quadratic order term arises due to ionization of \HI because of baryonic feedback. Note that the negative terms are only present in the $z=1$ case as feedback becomes stronger at low-$z$. For $z=5$, one can see that the response of $M_\textup{HI}$ is stronger for $\delta_{0.5}$ ($\alpha_{0.5}$) for the low (high) mass halos. Therefore, a combination of them will give a fairly monotonic trend, which is what we find in Eq.~\ref{eq:SyRegz5}.

It is also worth mentioning that our expressions in Eq.~\ref{eq:SyReg} do not capture all the scatter in the \HII-halo mass relation seen in Fig.~\ref{fig:HImassfn}. 
We have only modeled the part of the scatter connected to the environment and, as we will see in the next section, this is sufficient for improving the accuracy of the clustering of \HI by a significant amount.
See Ref.~\cite{TorVilinprep} for a more comprehensive modeling of \HII-halo mass scatter and its correlation with various baryonic properties of the halo.

\begin{figure*}
\centering
\includegraphics[scale=0.53,keepaspectratio=true]{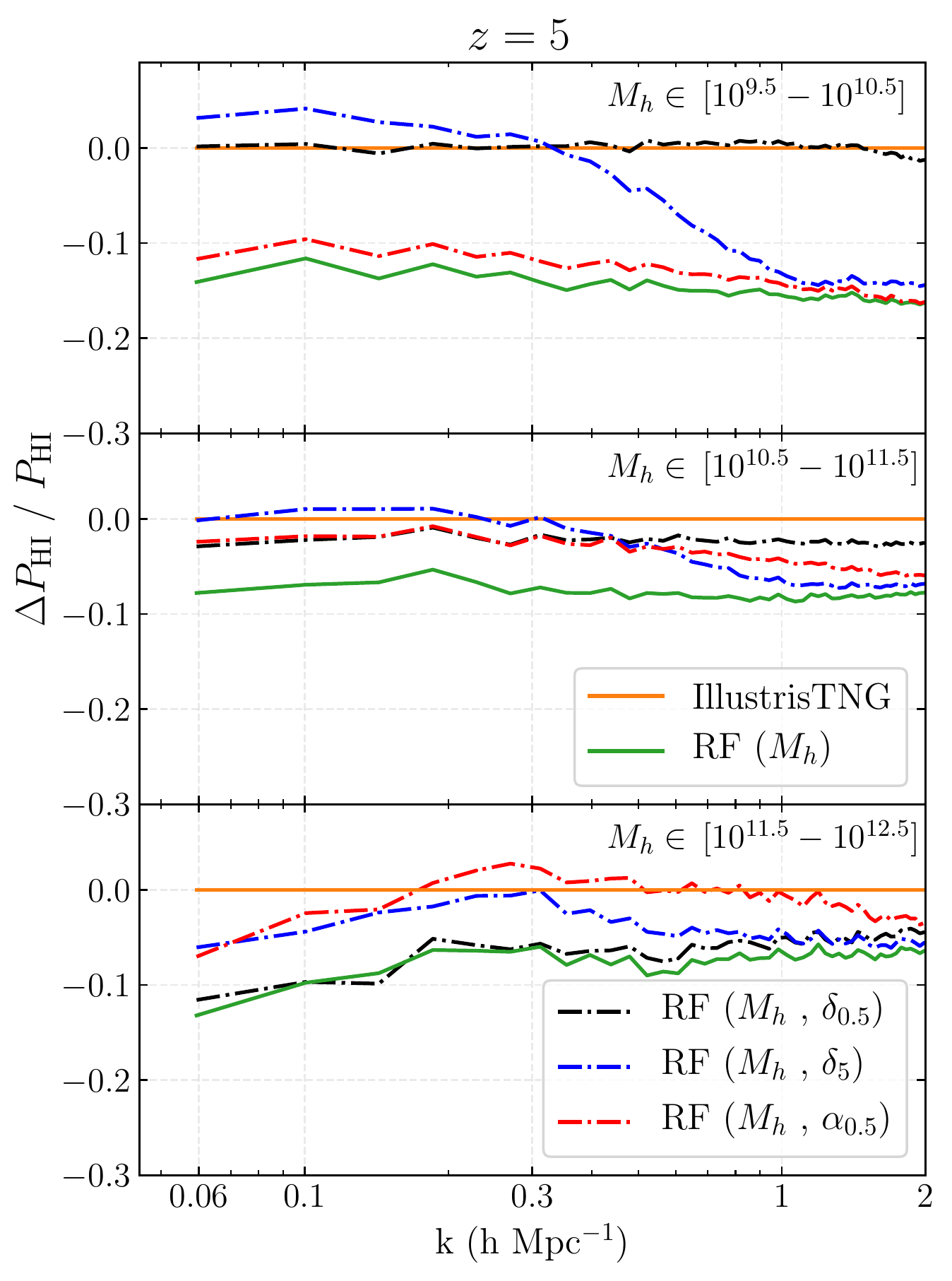}
\includegraphics[scale=0.53,keepaspectratio=true]{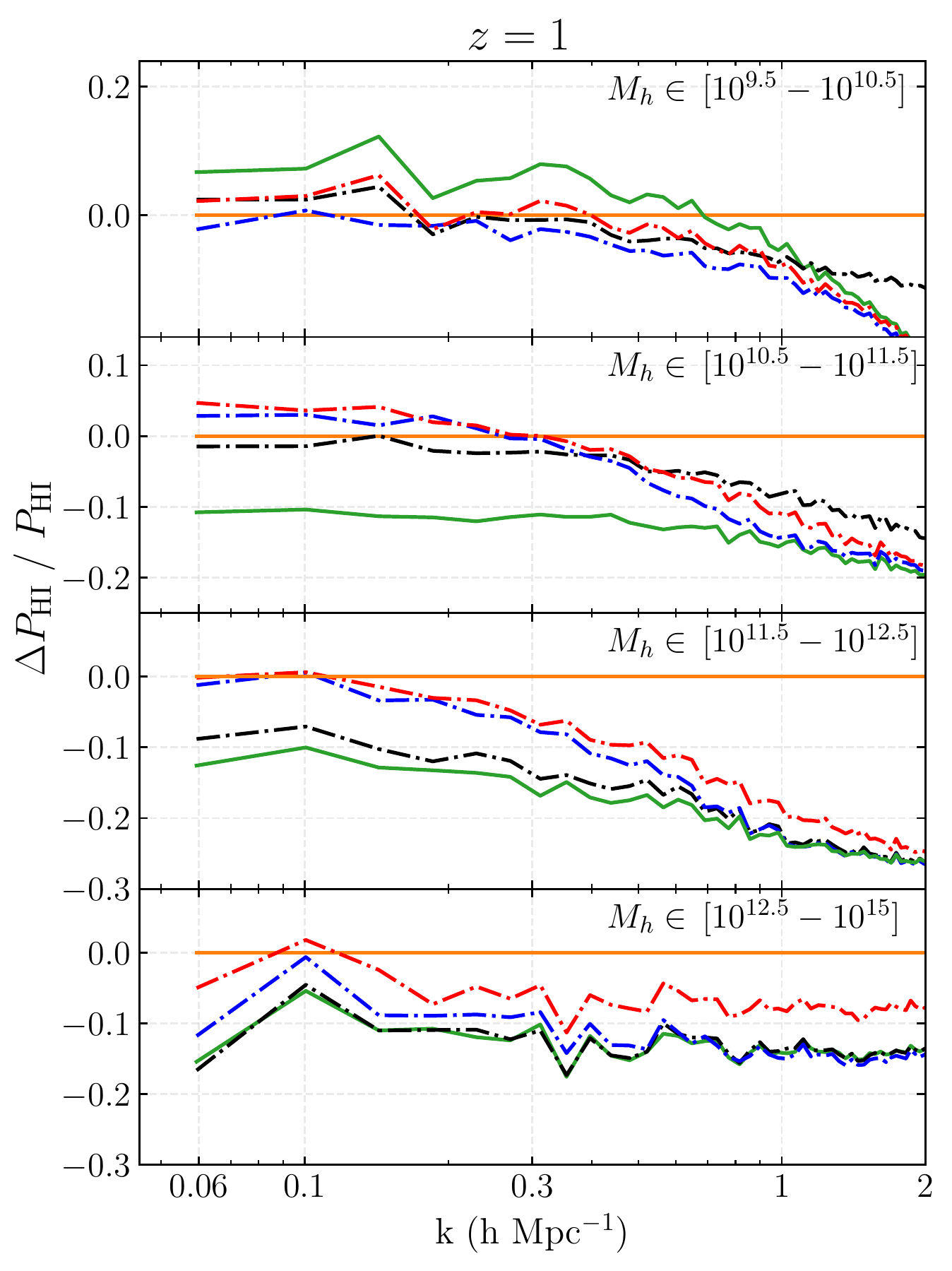}
\includegraphics[scale=0.53,keepaspectratio=true]{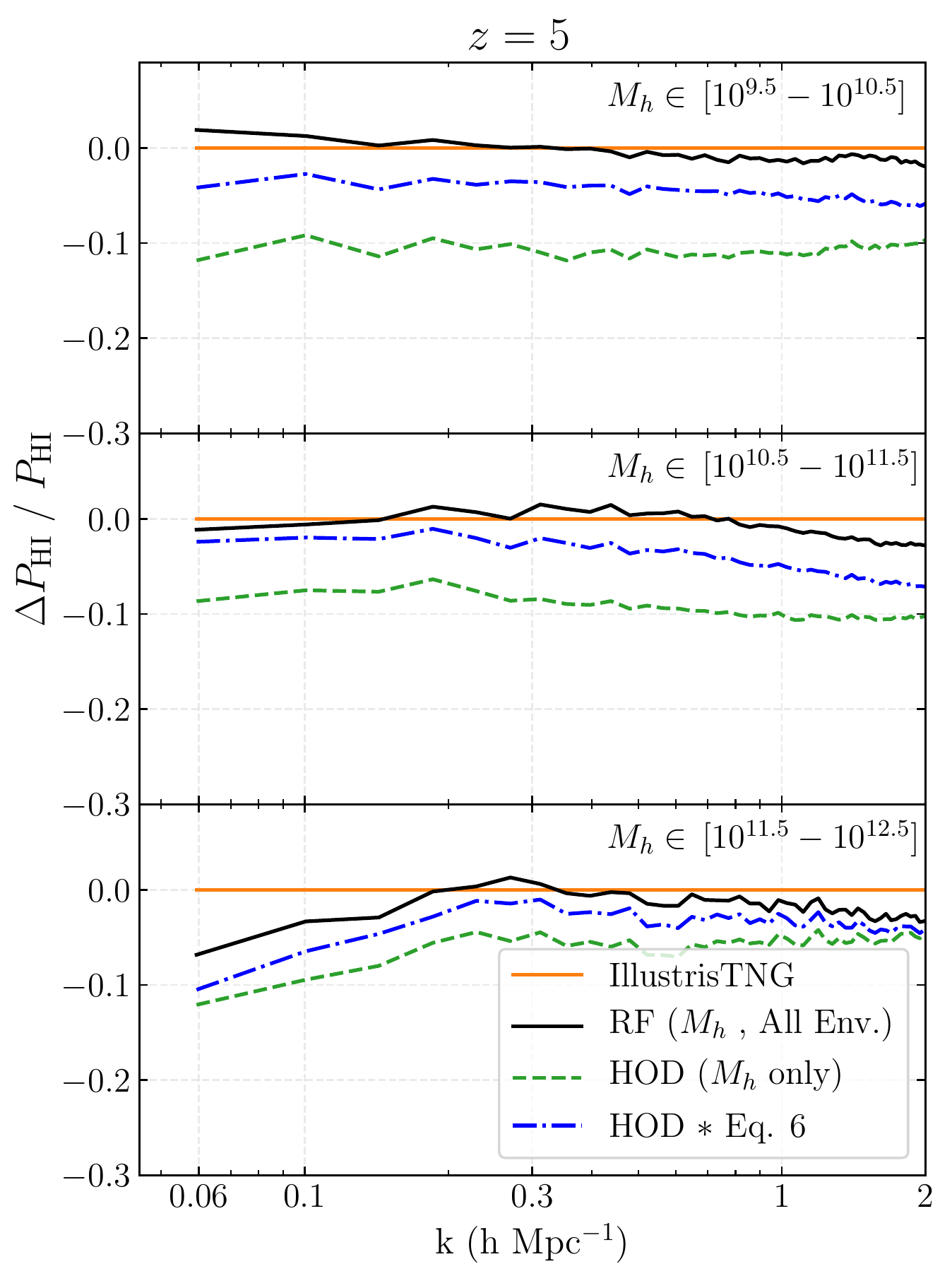}
\includegraphics[scale=0.53,keepaspectratio=true]{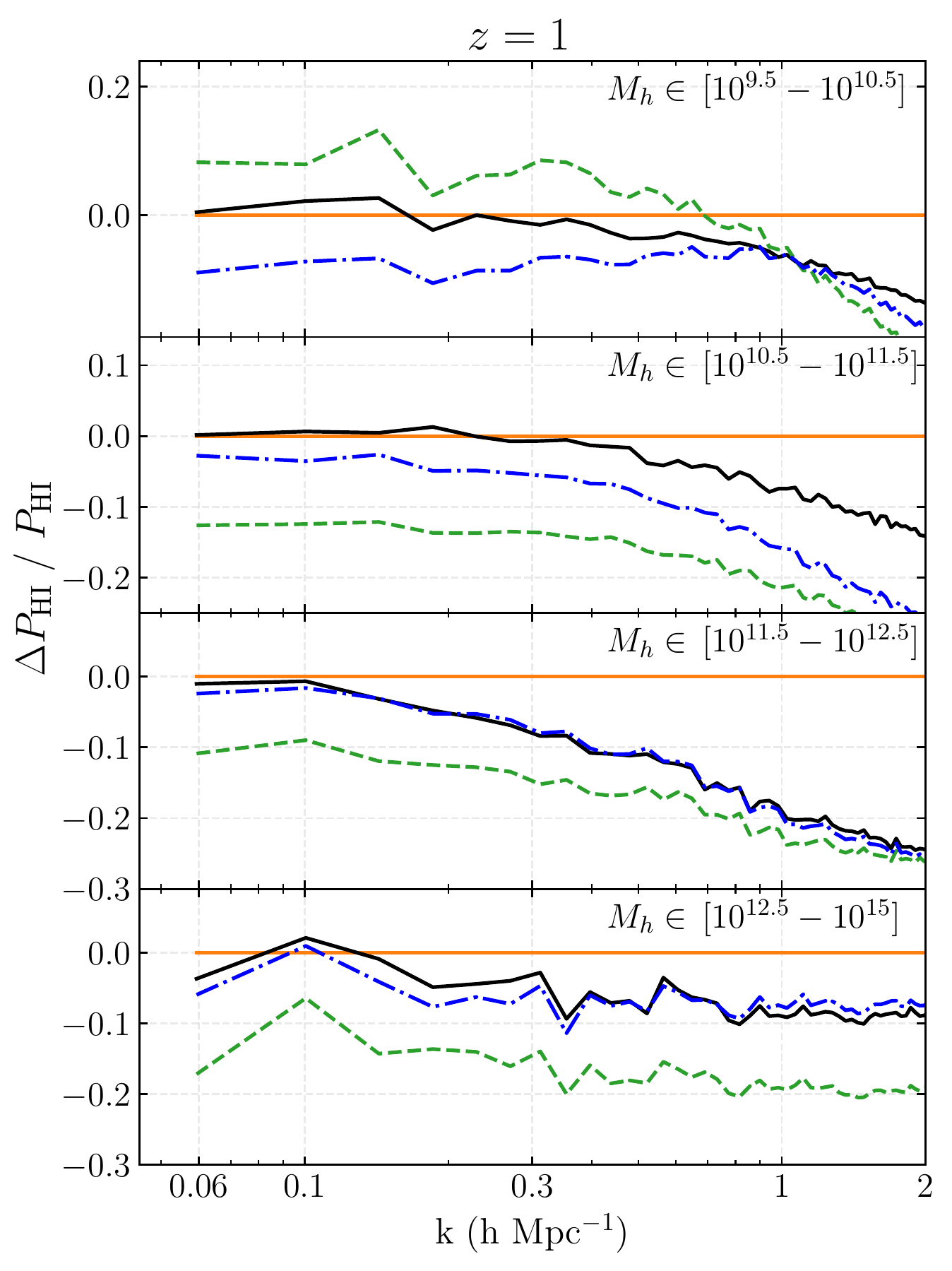}
\caption{This figure illustrates the effects of halo environment on the power spectrum of the modeled \HI field. We train random forest (RF) regressors to predict the \HI mass of halos as a function of different environmental properties (see legend).
The line legends are the same for the left and right panels.
\textbf{Top:} The RF is trained separately on three different environmental variables and the results show slightly different improvements in each case.
\textbf{Bottom:} The RF is trained using the environmental information (both the overdensity and tidal information) at various scales. We also show the result from the mass-only HOD model in dashed green which uses Eq.~\ref{eq:HOD} and the dot-dashed blue line shows the results from modeling the halo environment using Eq.~\ref{eq:SyReg}, which was derived using symbolic regression. Overall, we see that the predictions improve significantly for all halo masses when the environmental information is included in modeling of the halo \HI mass.
}
\label{fig:Pow_AllEnv}
\end{figure*}

\begin{figure}
\centering
\includegraphics[scale=0.55,keepaspectratio=true]{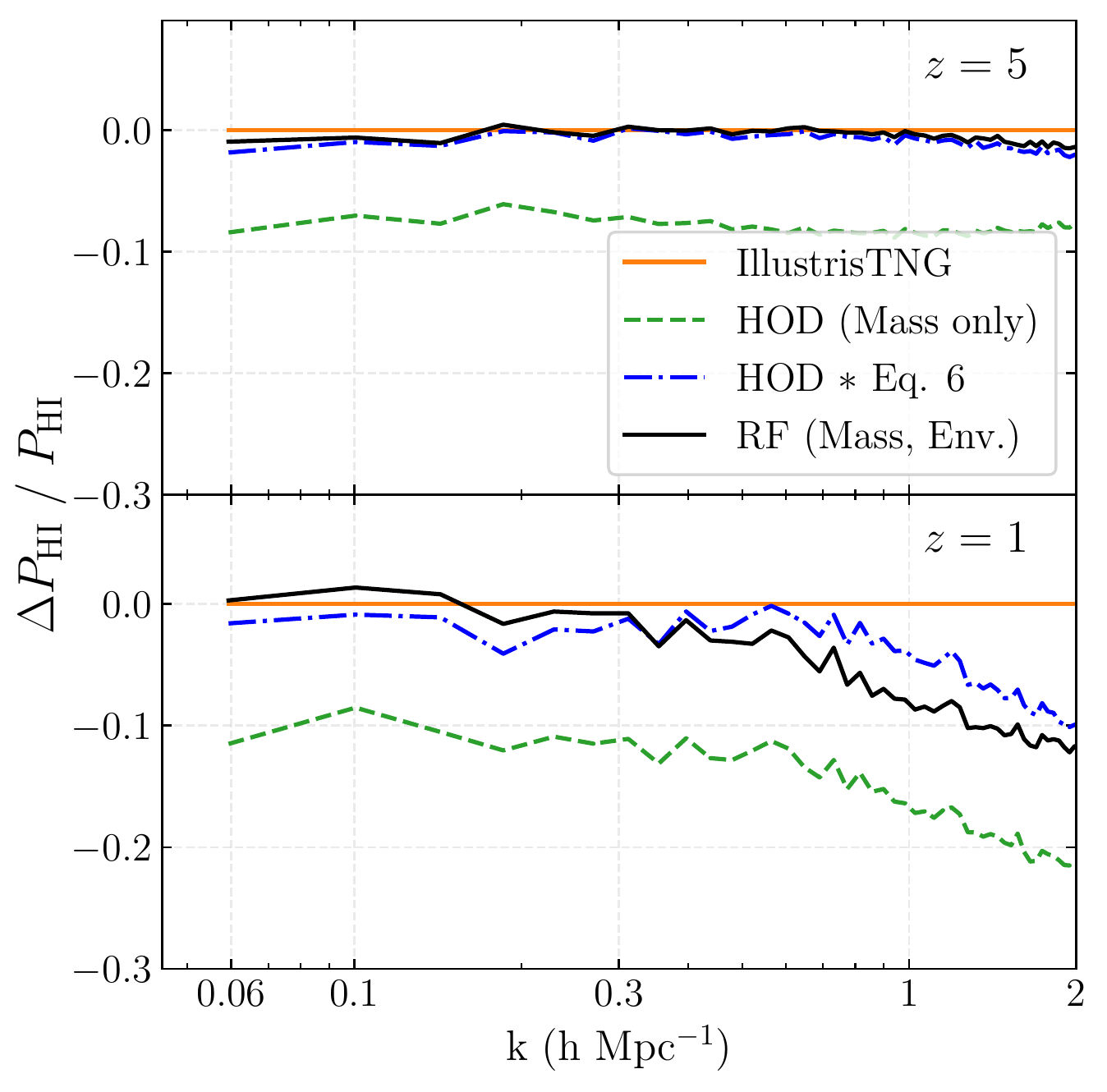}
\caption{Same as bottom panels of Fig.~\ref{fig:Pow_AllEnv} but for a combined set of halos of all masses. 
Note that modeling the effect of the halo environment using either random forests (RF) or symbolic regression significantly improves the accuracy of the predicted \HI field at both high and low redshifts; we observe a $\sim$ 10\% improvement in both cases. See Fig.~\ref{fig:OtherSummary} for a comparison of summary statistics other than $P_\textup{HI}$.}
\label{fig:SyReg_Pow}
\end{figure}

\section{Results for clustering of \HI}
\label{sec:results}
Till now we have discussed the effects of halo environment on its \HI mass. In this section, we investigate how this dependence propagates into the clustering of \HI (we refer to this phenomenon as \HI assembly bias in this paper). We will now focus on the real-space \HI power spectrum $P_\textup{HI}(k)$ as it is directly related to the 21-cm power spectrum (which will be measured from upcoming surveys via
$P_\textup{21cm}(k) = \bar{T}^2_b\, P_\textup{HI}(k)$, where $\bar{T}_b$ is the mean brightness temperature of the 21-cm line).
For results corresponding to other summary statistics like the bispectrum and cross-correlation coefficient, see Appendix~\ref{apx:OtherSummary}.

Let us first discuss an intuitive explanation of how the environment of the halos impacts the overall clustering of \HII.
It is well-known from excursion set theory that the environment of halos has a very strong impact on the clustering of halos themselves \cite{BonColEfs91} (for e.g., it is easier to form halos in a denser environment as the collapse threshold is more frequently reached and such halos therefore are significantly more clustered; this is indeed also the case in the TNG300-1 simulation as shown in Fig.~\ref{fig:PowSplit}). 
If halos in denser regions have even slightly more \HI than those in sparser regions, the overall clustering of \HI would increase; this is because the calculation of $P_\textup{HI}$ involves weighting halos by their \HI mass and the halos in denser regions would get upweighted.
We thus conclude that halo environment not only can affect the \HI content of a halo, but also the overall clustering signal of \HII.

In order to gauge the effect of various environmental variables on $P_\textup{HI}(k)$, we use a machine learning model called random forest regressor (RF). We first split the TNG300-1 box into training and test sets: the full box has a side-length $205 \Mpc$, from which we use a box with a side-length $150 \Mpc$ (comprising $\sim 40$\% of the total volume) for testing the RF and the rest for training the RF.
Using the halos in the training volume, we train the RF to predict the halo \HI mass as a function of various different halo environmental parameters,
and then use it to predict the \HI mass of halos in the test set.


We show the power spectrum of the generated \HI field from the RF in Fig.~\ref{fig:Pow_AllEnv}. For the top panels, we train the RF separately with three environmental parameters \{$\delta_{0.5}$,$\alpha_{0.5}$,$\delta_{5}$\}.
As discussed above, we see that including the environmental information causes the \HI clustering to increase in general (as halos in denser or more anisotropic environments typically have more \HI than average).
The only exception to this trend is for $M_h\in [10^{9.5}-10^{10.5}] \Ms$ at $z=1$. To understand this, we can look at this case in the lower panels of Fig.~\ref{fig:mHI} (blue line): as the environment becomes very dense, the \HI mass in the halos significantly decreases and therefore, on average, halos in less dense environments have more \HII.
We also see that a lot of the information from the environment at small-scales ($\delta_{0.5},\alpha_{0.5}$) is degenerate with the information from large scales ($\delta_5$).
 We again see that for $z=5$, $\delta_{0.5}$ and $\alpha_{0.5}$ give complementary trends, similar to our discussion on Eq.~\ref{eq:SyRegz1} earlier.
Another trend in Fig.~\ref{fig:Pow_AllEnv} worth noting is that including the variable $\delta_{5}$ does not improve clustering at small-scales ($k \gtrsim1\kMpc$), which is expected since halos at smaller separations are in the same large-scale environment. To compare with the environmental trends, we also show a case similar to the HOD, where the RF is only trained with the halo mass ($M_h$) (the RF is just predicting the mean of the scatter in Fig.~\ref{fig:HImassfn}). We do not show error-bars arising from cosmic variance because all the cases are evolved from the same initial conditions. 
 We have only shown the effect of environmental variables in Fig.~\ref{fig:Pow_AllEnv} and show the effect of internal properties of the halo on $P_\textup{HI} (k)$ in Fig.~\ref{fig:Pow_Others}.

Instead of training the RF with a single environmental parameter, we now use all the environmental overdensity and anisotropy parameters at the scales \{0.5,1,2,5,10,20\} $\Mpc$, and show the corresponding results in the bottom panel of Fig.~\ref{fig:Pow_AllEnv}. We also show results obtained by using Eq.~\ref{eq:SyReg} and we see that the relative improvement is smaller compared to RF. It is important to note that this is because the RF was trained separately for each of the four mass-bins in the figure, while the symbolic regressor was trained on a combined sample that included all halo masses---if we train the symbolic regressor for each individual mass bin, we expect to see better results. Finally, we make a combined sample comprising of halos from all mass bins and show the corresponding results in Fig.~\ref{fig:SyReg_Pow}.
Indeed, as the symbolic regressor was trained for the combined sample,
it shows better results as compared to Fig.~\ref{fig:Pow_AllEnv}, and the improvement due to Eq.~\ref{eq:SyReg} and RF is comparable.


\section{Discussion and Conclusions}
\label{sec:Conclusions}
Upcoming galaxy and line intensity mapping surveys will map large volumes of the Universe.  We need accurate large-scale mock baryonic catalogues to provide the theory predictions needed to maximize the scientific return of these missions. Halo model tools like HOD are widely used for making large-scale baryonic maps and typically make the assumption that the baryonic content of a halo is a function of only the halo mass.


Using the IllustrisTNG simulation, we show that the neutral hydrogen (\HII) content of a halo is dependent on secondary properties other than halo mass, like the environment of the halo, the mass fraction of substructure inside the halo and its concentration (see Figs.~\ref{fig:mHI} and~\ref{fig:mHI2}). We show that these secondary dependences also affect the overall clustering of \HI and lead to \HI assembly bias. We also provide physical explanations for the dependences, and
model the effect of the halo environment on its \HI mass using machine learning tools like random forests and symbolic regression.
Our modeling can be easily used to augment a mass-based HOD model of \HII, and leads to a significant improvement in the clustering of the modeled \HI field (the real-space 21-cm power spectrum prediction is improved by $\gtrsim$10\% on scales $k\gtrsim 0.05 \kMpc$, see Fig.~\ref{fig:SyReg_Pow}).
Modeling the assembly bias effects using parameters related to the halo environment has the additional advantage that these parameters can be easily computed in DM-only simulations, without there being a need to construct halo merger trees or resolve sub-halo structure. 

In order to appropriately marginalize over assembly bias 
in a Bayesian analysis of survey data, it is crucial to encode its effects in a compact analytic expression. 
Symbolic regression enables such an encoding (see Eq.~\ref{eq:SyReg}) and is therefore more advantageous to use over machine learning techniques like random forests or neural networks. Furthermore, the results from symbolic regression can provide an understanding of the underlying physical behavior and are readily generalizable.
We expect symbolic regression to be an ideal tool for parameterizing assembly bias for any general case of baryonic tracers, directly by using data from hydrodynamic simulations or semi-analytic models.


\subsection*{Comparison with other works on modeling the halo environment}
Although our study focuses on \HII, it is worth comparing our approach of modeling the environment to that of other studies which modify the HOD formalism for galaxies based on the halo environment. 
There has been an ample interest on including the effect of the environment of the halo into the standard HOD model for populating galaxies \cite{McEWei18, XuZeh20, WibSalWei19, SalWibWei20,YuaHadBos20,HadTac20,SalZu20} in order to create more accurate galaxy mock catalogs. 
The previous studies however have only used a single parameter for modeling the halo environment, and they rely on the trends being monotonic with respect to that chosen parameter. As seen in some cases in Fig.~\ref{fig:mHI}, the trends can have turnovers and be non-monotonic for parameters like the environmental overdensity. In such cases studying the effect of a single parameter could give misleading results (for e.g., the increasing and the decreasing part of the trends can cancel out giving a null result overall). 
Furthermore, it is optimal at times to use a combination of two parameters to model the assembly bias effect (as seen in Eq.~\ref{eq:SyRegz5} where $\delta_{0.5}$ and $\alpha_{0.5}$ are both used). 
Symbolic regression is therefore an alternative approach to infer an optimal and physically motivated parameterization of assembly bias directly from simulations.

\subsection*{Future work} 
There are multiple ways in which our work can be extended. We have used the results from the IllustrisTNG simulation which uses a particular prescription for baryonic feedback.
We plan to see how our results---in particular, the structure of Eq.~\ref{eq:SyReg}---depend on astrophysical feedback parameters using the CAMELS suite of simulations \cite{VilAngGen20} (which contain 2184 hydrodynamic simulations run for different astrophysical feedback and cosmology parameters).
Such a test will also be useful in studying how the measurements of clustering of \HI from upcoming surveys can be used to constrain feedback prescriptions in hydrodynamic simulations. We have performed our analysis for two particular redshifts ($z=$\{1,5\}) separately, and we plan to analyze intermediate redshifts and derive a single equation similar to the ones in Eq.~\ref{eq:SyReg}, but which also includes a redshift dependence. We have studied the assembly bias effects in real space and it will be interesting to extend our analysis to redshift space and probe potential anisotropic assembly bias effects because of the correlation between the local tidal field and the \HI mass of the halo.
We will also extend our techniques from \HI to galaxies (model the number of galaxies in a halo instead of the halo \HI mass) in an upcoming paper.




 

\acknow{We thank Hamsa Padmanabhan, Boryana Hadzhiyska, Sownak Bose, Daniel Eisenstein, Neal Dalal, Sandy Huan, Roman Scoccimarro, Chang Hahn, Andrej Obuljen, Jeremy Tinker and Miles Cranmer for fruitful discussions. We also thank Hamsa Padmanabhan for detailed comments on the draft of this paper.
FVN acknowledges funding from the WFIRST program through NNG26PJ30C and NNN12AA01C. The work of SH is supported by Center for Computational Astrophysics of the Flatiron Institute in New York City. The Flatiron Institute is supported by the Simons Foundation. 
This work was also supported in part through the NYU IT High Performance Computing resources.
We thank the IllustrisTNG collaboration for making their simulation data publicly available.
We have used the publicly available Pylians3 libraries\footnote{\url{https://github.com/franciscovillaescusa/Pylians3}} to carry out the analysis of the simulations and \textsc{PySR}$^{\ref{PySR}}$ package for symbolic regression.}

\showacknow{} 


\clearpage
\appendix
\setcounter{equation}{0}
 \setcounter{table}{0}
 \setcounter{figure}{0}

 \renewcommand{\theequation}{S\arabic{equation}}
 \renewcommand{\thefigure}{S\arabic{figure}}
 \renewcommand{\thetable}{S\arabic{table}}
 
\begin{figure*}
\begin{center}
{\LARGE \bf Supplemental material for `Modeling the neutral hydrogen assembly bias with machine learning and symbolic regression'}
\end{center}
\end{figure*}

\begin{figure*}
\centering
\includegraphics[scale=0.42,keepaspectratio=true]{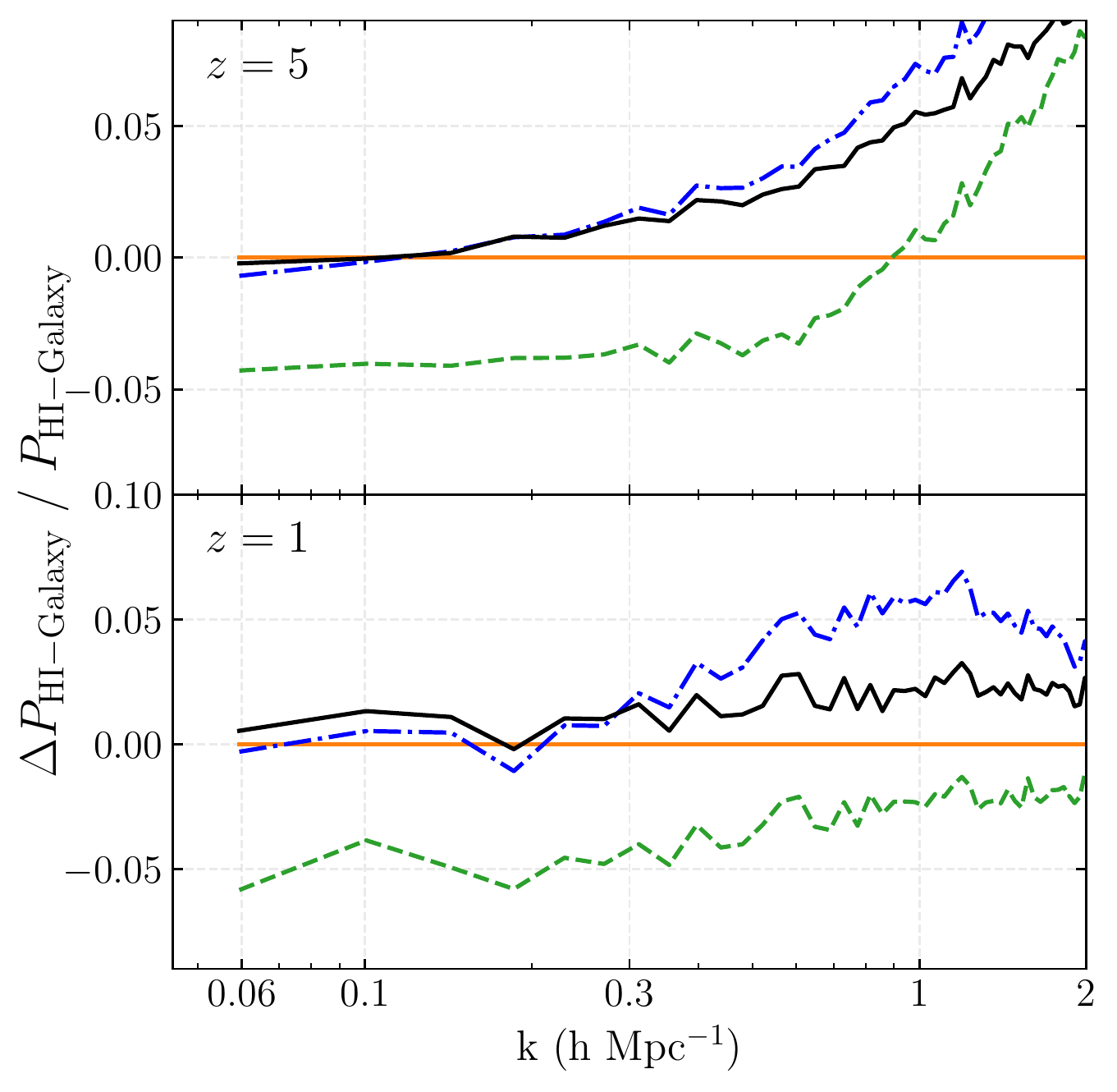}
\includegraphics[scale=0.42,keepaspectratio=true]{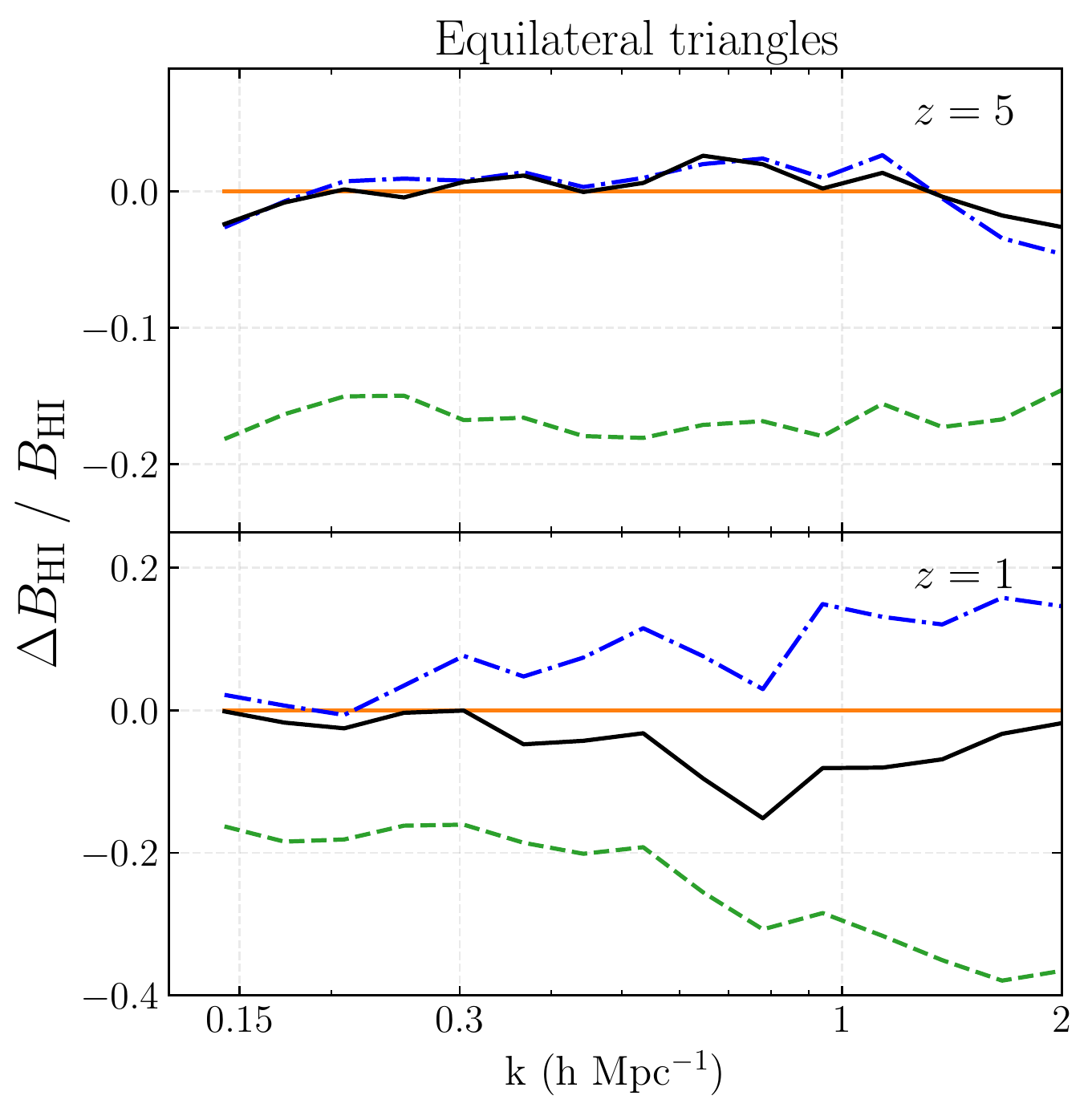}
\includegraphics[scale=0.42,keepaspectratio=true]{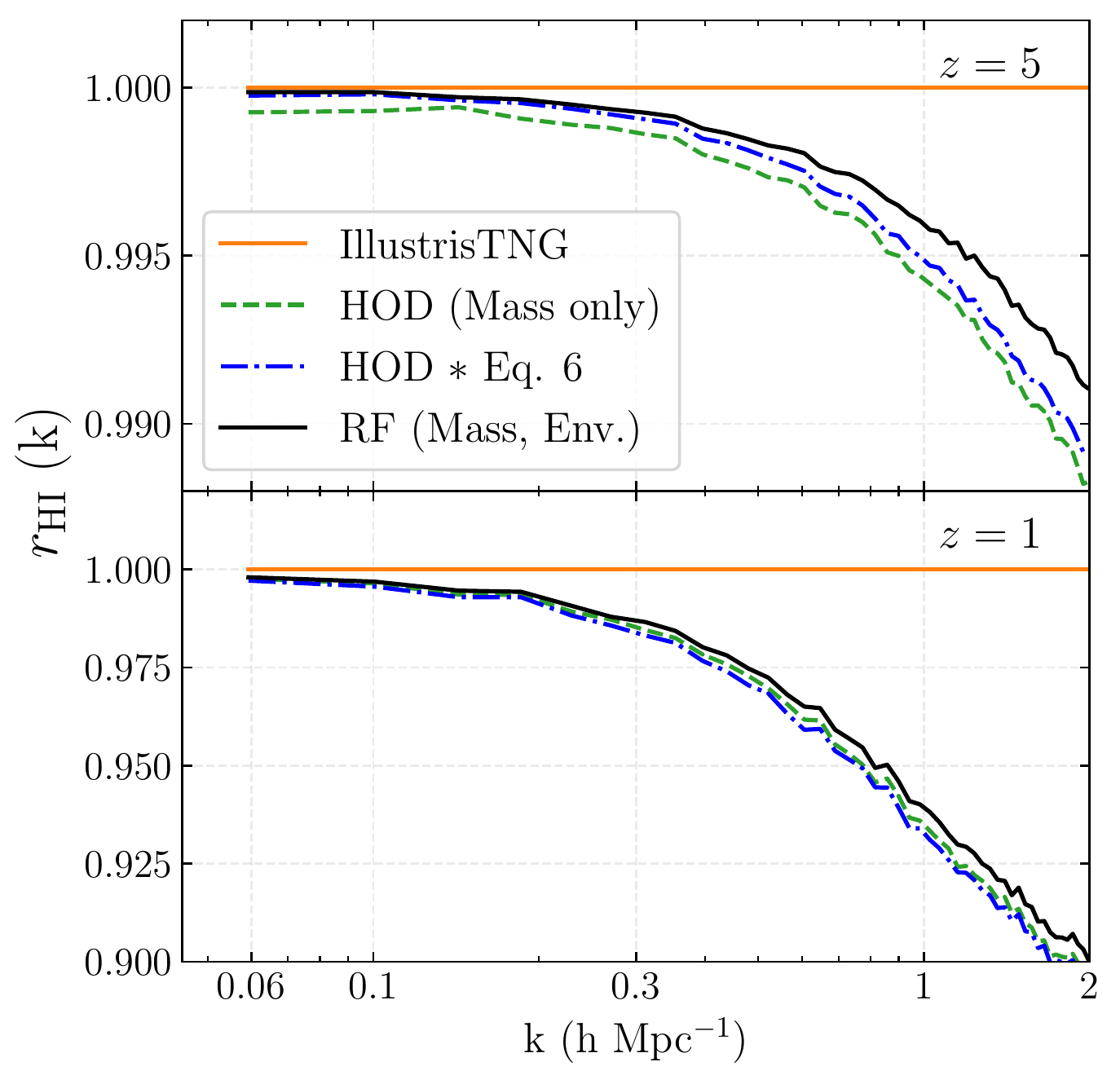}
\caption{Same as Fig.~\ref{fig:SyReg_Pow}, but for different summary statistics of the modeled \HI field. \textbf{Left:} Cross power between \HI and galaxies, where we use galaxies in the TNG300-1 box with $M_*>10^{10}M_\odot$. Note that the deviation in the cross power at high-$k$ at $z=5$ is due to low number density of galaxies (we have checked that increasing the number density improves our results at high-$k$ for both redshifts). \textbf{Center:} Bispectrum as a function of side length for the equilateral triangle configuration. \textbf{Right:} The cross-correlation coefficient of the modeled \HI field (defined as $r_\mathrm{(A)}= P_\mathrm{A-Illustris}/\sqrt{P_\mathrm{A}P_\mathrm{Illustris}}$). Overall, including the halo environment provides significant improvement for $P_\mathrm{HI-Galaxy}$ and $B_\textup{HI}$ but a modest change in $r_\textup{HI}$. This hints at the fact that the halo environment primarily affects the bias of the modeled \HI field.}
\label{fig:OtherSummary}
\end{figure*}

\section{Other summary statistics of the modeled \HI field}
\label{apx:OtherSummary}
In Sec.~\ref{sec:results}, we showed the comparison between the auto-power spectrum of the \HI field modeled via HOD, random forests (RF) and symbolic regression. In this section, we discuss other useful summary statistics of the modeled \HI field and show the corresponding results in Fig.~\ref{fig:OtherSummary}. 

\subsubsection*{1) Cross-power between HI and galaxy fields}
$P_{\textup{HI} - \textup{Galaxy}}$ is an important statistic to study because large regions of future \HI surveys will overlap with those of galaxy surveys like DESI or the Roman space telescope. Furthermore, unlike the auto-power spectrum of 21cm which is yet to be detected, there have been multiple detections of the $P_{\textup{HI} - \textup{Galaxy}}$ signal at $z\sim1$ \cite{ChaPen10, MasSwi13}.
To calculate $P_{\textup{HI} - \textup{Galaxy}}$, we use galaxies in the TNG300-1 sample with stellar masses M$_*> 1\times10^{10}$ M$_\odot$
(which corresponds to a number density of $n=2\times10^{-4}$ h$^3$/Mpc$^3$ at $z=5$ and $n=9\times10^{-3}$ h$^3$/Mpc$^3$ at $z=1$). 
We see that, similarly to $P_\textup{HI}(k)$ in Fig.~\ref{fig:SyReg_Pow}, modeling the halo environment improves the $P_{\textup{HI} - \textup{Galaxy}}$ prediction in the left panel of Fig.~\ref{fig:OtherSummary}. It is worth noting that there is a deviation for all modeling techniques at high-$k$ at $z=5$.
We have checked that increasing the number density of the galaxy sample removes this high-$k$ deviation. This suggests that the deviation is a result of the large shot noise present in the $z=5$ galaxy sample. We emphasize that we have not included the one halo term in our analysis and including it will likely improve the predictions for $k\gtrsim1 \kMpc$ for all summary statistics.


\subsubsection*{2) Bispectrum}
Late-time gravitational clustering causes a significant leakage of cosmological information, that initial was in the power spectrum, into higher order statistics \citep{WadSco19,ScoZalHui9912,TakJai04,Quijote,Chang_Bk}. To recover this information, the lowest order statistic that one needs to compute in Fourier space is the bispectrum. Unlike the power spectrum, the bispectrum is sensitive to the shape of structures generated by late-time gravitational instability, and therefore provides complementary information.
We show results for the bispectrum in the center panel of Fig.~\ref{fig:OtherSummary}. The HOD model again shows a deviation, similar to the power spectrum case, at low-$k$ and including the environmental effects improve the prediction. We only show results for the equilateral triangle configuration but have checked that the improvement is similar for other triangle configurations.


\subsubsection*{3) HI cross-correlation coefficient} The auto-power spectrum $P_\textup{HI}$ measures the amplitude of \HI fluctuations (averaging over the squares of the mode amplitudes), being thus insensitive to fluctuations phases. We therefore calculate the cross-correlation coefficient of the \HI field (which is sensitive to the phases of \HI fluctuations), and show the results in the right panel of Fig.~\ref{fig:OtherSummary}. 
We see that including the environmental effects provides marginal improvement at $z=5$ and has a negligible effect at $z=1$. Unlike the other statistics we discussed,  the cross-correlation coefficient is not sensitive to the bias of \HII. We therefore conclude that the modeling the environment of the halo improves the prediction of the \HI bias, but has a small effect on the phases of the \HI fluctuations.

\section{Effect of internal halo properties on its \HI mass}
\label{apx:OtherParams}

In Sec.~\ref{sec:Env_mHI}, we discussed the effect of the environment properties of the halo on its \HI mass. Here we discuss the effect of two internal properties: halo concentration and subhalo mass fraction. 
Apart from these two, there could also be correlations between $\mHI$ and other internal halo properties like its formation epoch, spin, velocity dispersion and others, but we leave their study to a future work.

\subsubsection*{Mass fraction in subhalos}
As shown earlier in Fig.~\ref{fig:mHI2}, the fractional mass in subhalos is related to the total \HI inside a halo.
This is because the gas in subhalos is more dense and can more efficiently self-shield itself as compared to gas in the CGM (circumgalactic medium) of the halo. The subhalo mass fraction is correlated with the environment of the halo (increasing the environmental overdensity typically results in more mergers and therefore more substructure in halos). We show the magnitude of this correlation in Fig.~\ref{fig:mHI_others};
we have used an intermediate length scale (2.5 $\Mpc$) to showcase the trends (we also checked that we obtain similar results using some other environmental parameters, like $\alpha_{0.5}$). Note that we use the subhalo data provided by IllustrisTNG in our analysis \cite{SprWhi01}.

In order to show that the subhalo mass fraction also affects the clustering of \HII, we train a random forest regressor using it and show the results in Fig.~\ref{fig:Pow_Others}. We find improvements in the prediction of $P_\textup{HI}$ in most of the cases. 
There is however one particular case which is an exception: the low mass halos in $z=1$, where ram pressure stripping has the largest effect. 
Ram pressure stripping exclusively affects the gas distribution and not the DM distribution in the halo. Information of the subhalo mass fraction does not therefore capture this effect, leading to slightly biased predictions.


\subsubsection*{Halo concentration}
We use the following formula as a proxy to estimate halo concentration, $c$: $c=R_\textup{200c}/R_\textup{max}$, where $R_\textup{max}$ is the comoving radius at the point where the maximum circular velocity is attained for the largest subhalo inside the halo.
Note that this definition is less accurate for large mass halos; a more accurate way of finding the halo concentration is to fit a NFW profile to the halo mass distribution and measure the corresponding scale radius $r_s$ (the concentration can then be calculated as $c=R_\textup{200c}/r_s$). We have used the former manner to estimate concentration as the public IllustrisTNG data only provides information about $R_\textup{max}$.

One might expect the concentration of a halo to affect its \HI mass, as a deeper gravitational potential well can lead to larger amounts of gas being accumulated at the center (which helps in shielding from the UV radiation). This is however only true for the low-mass halos, as seen in the right panel of Fig.~\ref{fig:mHI2}. The situation becomes complicated for larger halos because higher concentrations lead to increased star formation and larger ionizing feedback due to supernovae and AGN. We indeed see this effect in the right panel of Fig.~\ref{fig:mHI2} where, beyond a certain threshold, larger concentration leads to lower $M_\textup{HI}$.
Upon training the RF with the halo concentration, we find very little improvement in the \HI clustering prediction. 

\begin{figure*}
\centering
\includegraphics[scale=0.55,keepaspectratio=true]{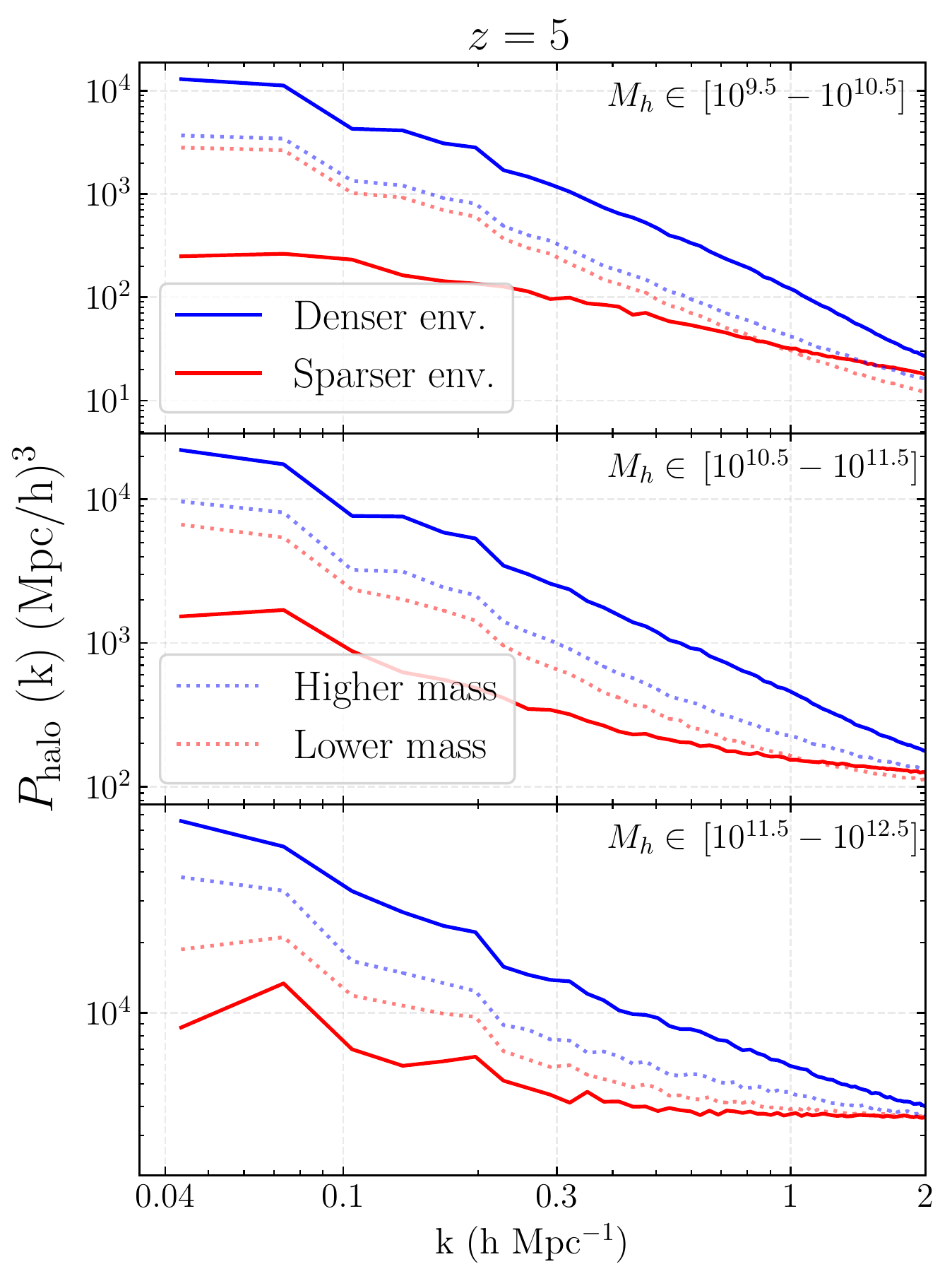}
\includegraphics[scale=0.55,keepaspectratio=true]{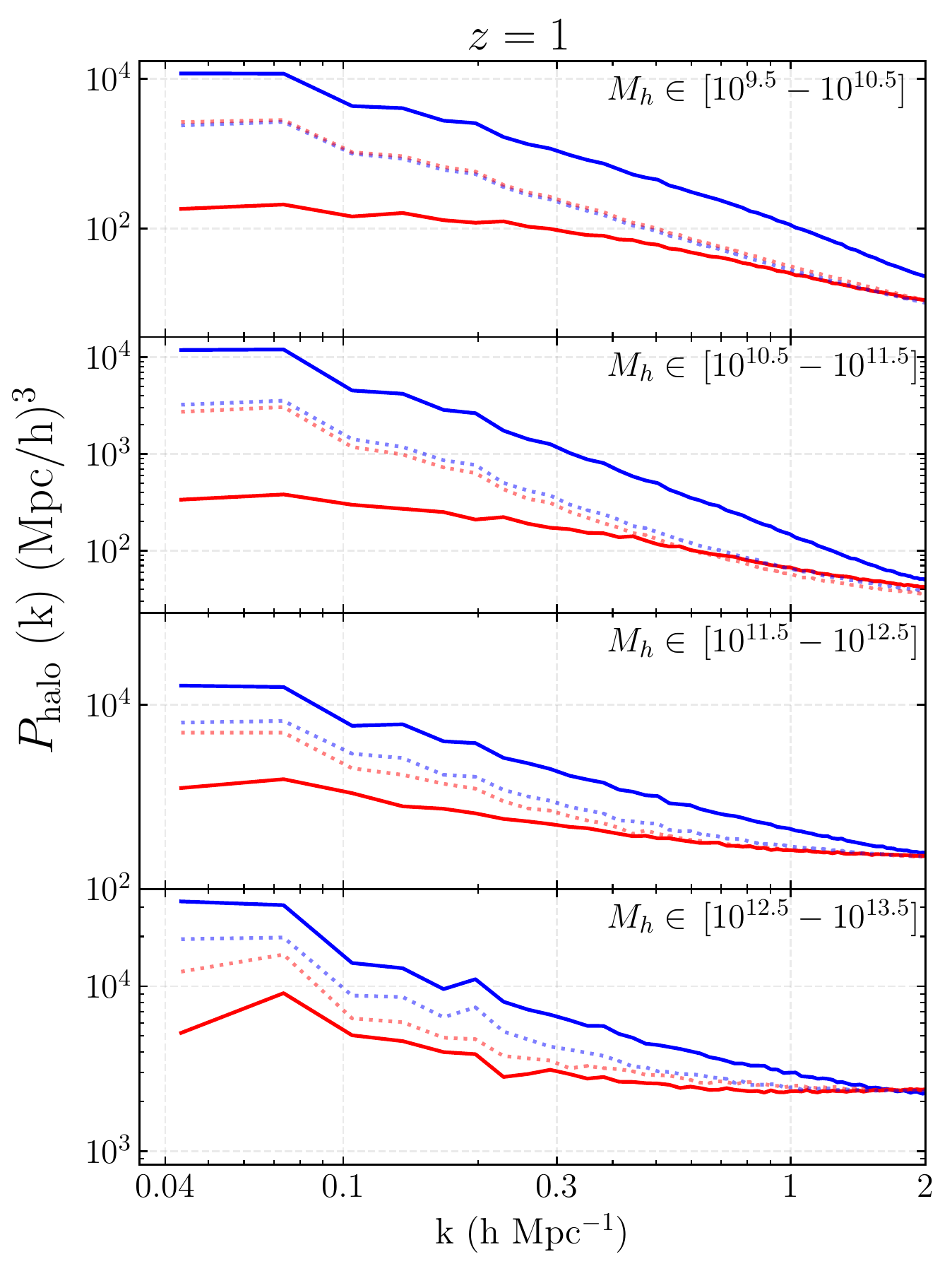}
\caption{Difference in the clustering of halos in denser and sparser environments for different halo masses ($M_h$). We split the halos in each mass bin into two sets based on their environmental overdensity: those with $\delta_{2.5}$ values above (Denser env.) or below (Sparser env.) the median; we show the halo power spectrum for the two samples in solid lines.
The halos in denser environments cluster a lot more strongly. Even if there is slightly more \HI than average in halos in denser environments (which amounts to upweighting their contribution to the power spectrum of \HII), the overall \HI clustering will increase. To confirm that the differences observed in halo clustering are not just due to variation in halo mass within our mass bins, we show in dotted lines the results from splits based purely on the halo mass.
All masses are in units of $\Ms$.}
\label{fig:PowSplit}
\end{figure*}

\section{Calibrating the HOD model for halos from $N$-body simulations}
\label{apx:Halos_FPvsDMO}

In Sec.~\ref{sec:HOD}, we discussed a mass based HOD model for \HII. In order to calibrate the model parameters $\{M_0,\alpha,M_\textup{min}\}$ in Eq~\ref{eq:HOD}., we had used the data for the total mass $M_h$ and the \HI mass $M_\textup{HI}$ of halos 
 from the hydrodynamic IllustrisTNG simulation. However, if this calibrated model is directly used on halos in an $N$-body simulation, one would get spurious results. This is because there is a significant change in the halo mass when baryons and their feedback are included in the simulation, as is seen in multiple hydrodynamic simulations \cite{LovPilGen18, SchFre15}. To quantify this change, we take advantage of the fact that TNG provides both the hydrodynamical (or full-physics, FP) simulation output as well as the $N$-body (or dark matter only, DMO) one, evolved from the same set of initial conditions. TNG also provides the matches for subhalos in the FP and DMO simulation (based upon their origin from the same Lagrangian patch of initial conditions). This helps us to match the corresponding halos in the FP and DMO simulations (following \cite{HadBosEis20}, we match two halos if their central subhaloes are matched).

\begin{figure*}
\includegraphics[scale=0.42,keepaspectratio=true]{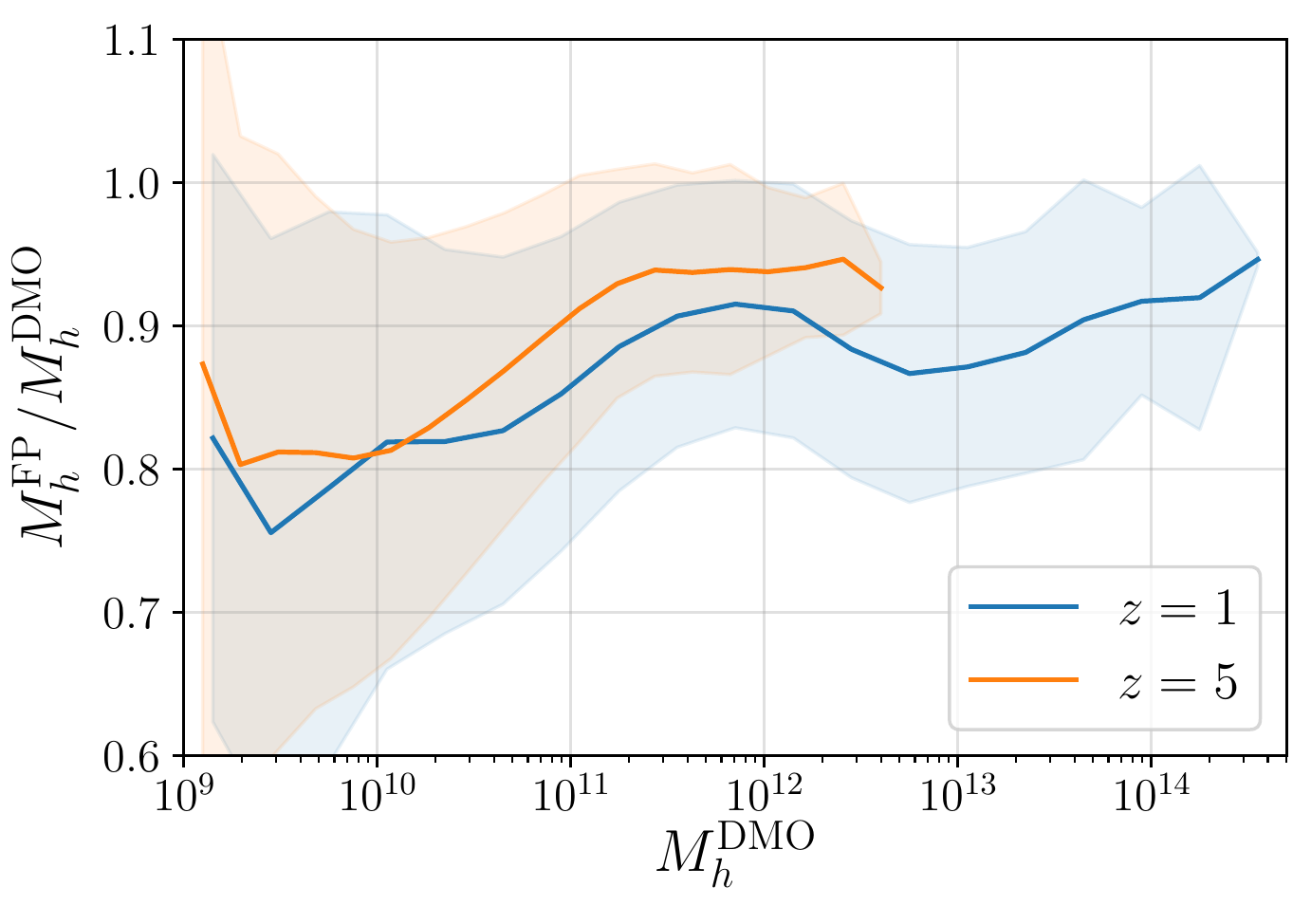}
\includegraphics[scale=0.42,keepaspectratio=true]{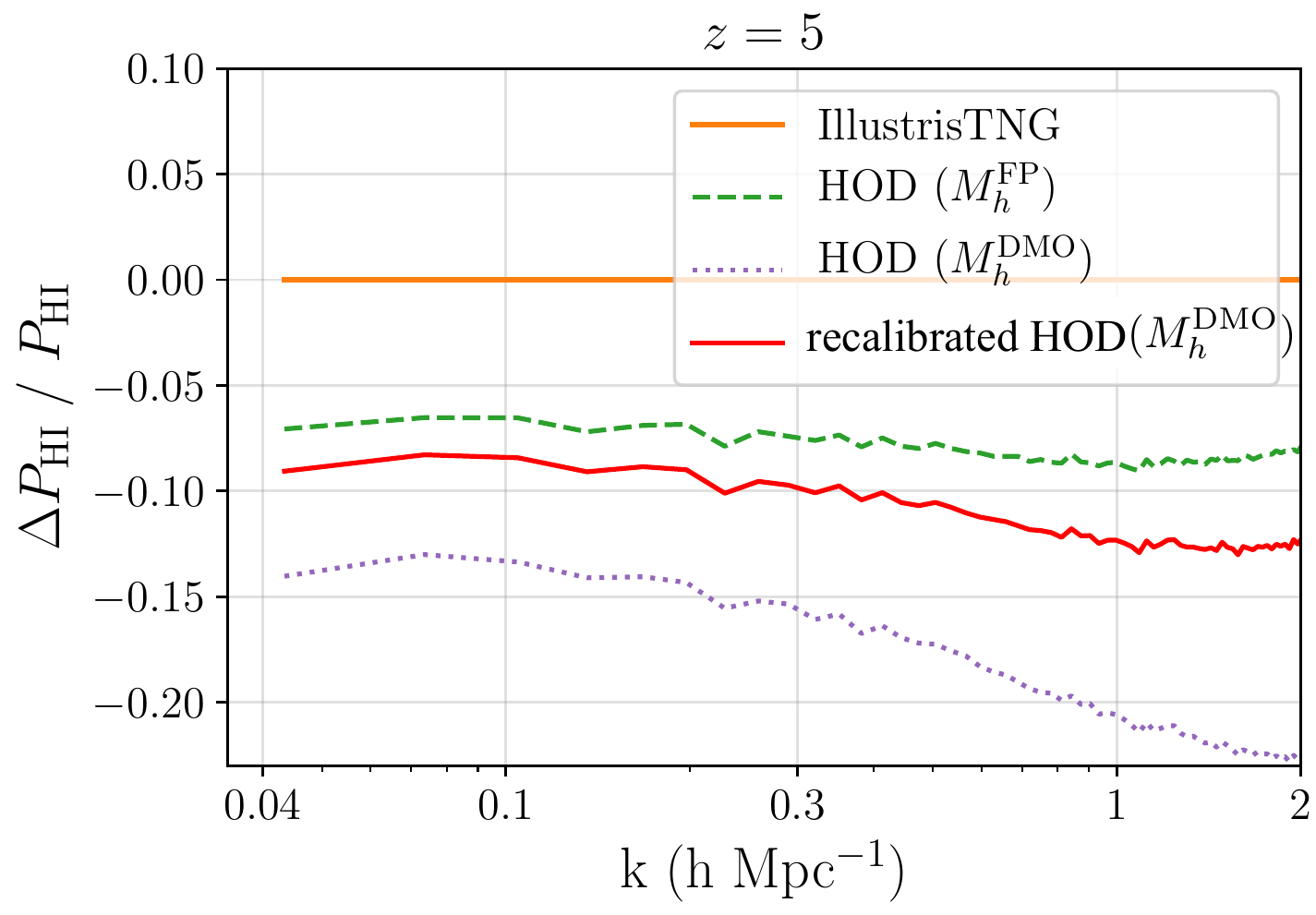}
\includegraphics[scale=0.42,keepaspectratio=true]{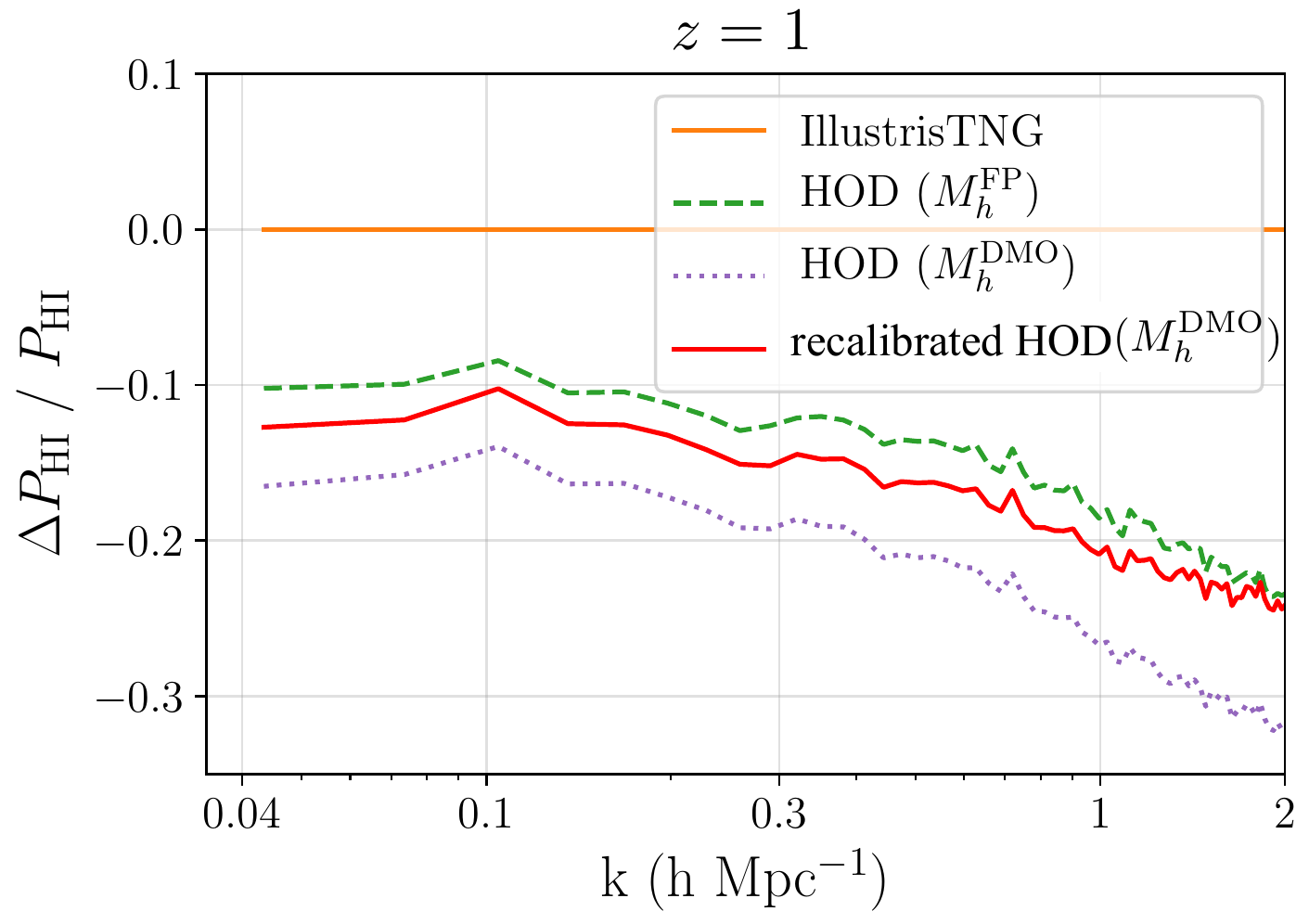}
\caption{This figure illustrates that an HOD relation, which is calibrated using halo masses from a hydrodynamic simulation, will give discrepant results if directly used on halos in a $N$-body simulation. We first match the halos in the hydrodynamic (full physics, FP) simulation to their counterparts in the $N$-body (DM-only, DMO) simulation.
\textbf{Left:} The ratio of the halo masses in the two simulations (the mean is in solid and the shaded area encloses 68 \% of the data). There is a significant difference in halo masses in the two simulations due to the effects of baryonic feedback.
\textbf{Center:} The relative difference in $P_\textup{HI}$ for the \HI field modeled using the HOD relation on halos from the FP (DMO) simulation is shown in green (purple). Note that the HOD relation was calibrated on data from the FP simulation (see Fig.~\ref{fig:HImassfn}), and therefore underperforms when used on the halos from the DMO simulation. We recalibrate the HOD model using halo masses from the DMO simulation and show the results in red, where the performance of the HOD model is now improved.
\textbf{Right:} Same as the center panel but for $z=1$.}
    \label{fig:DMOvsFP}
\end{figure*}

\begin{figure*}[p]
\vspace{-0.8cm}
\centering
\includegraphics[scale=0.53,keepaspectratio=true]{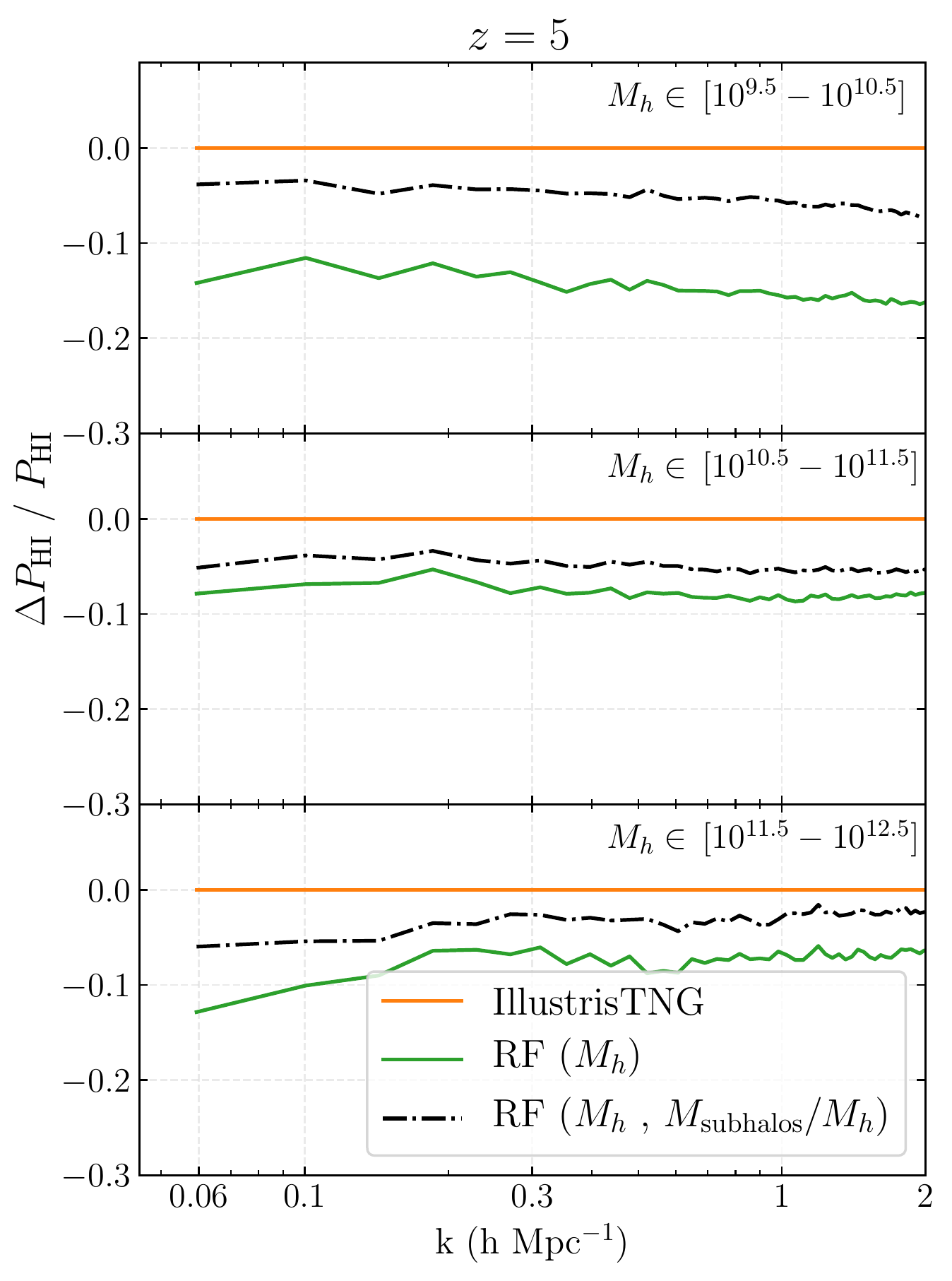}
\includegraphics[scale=0.53,keepaspectratio=true]{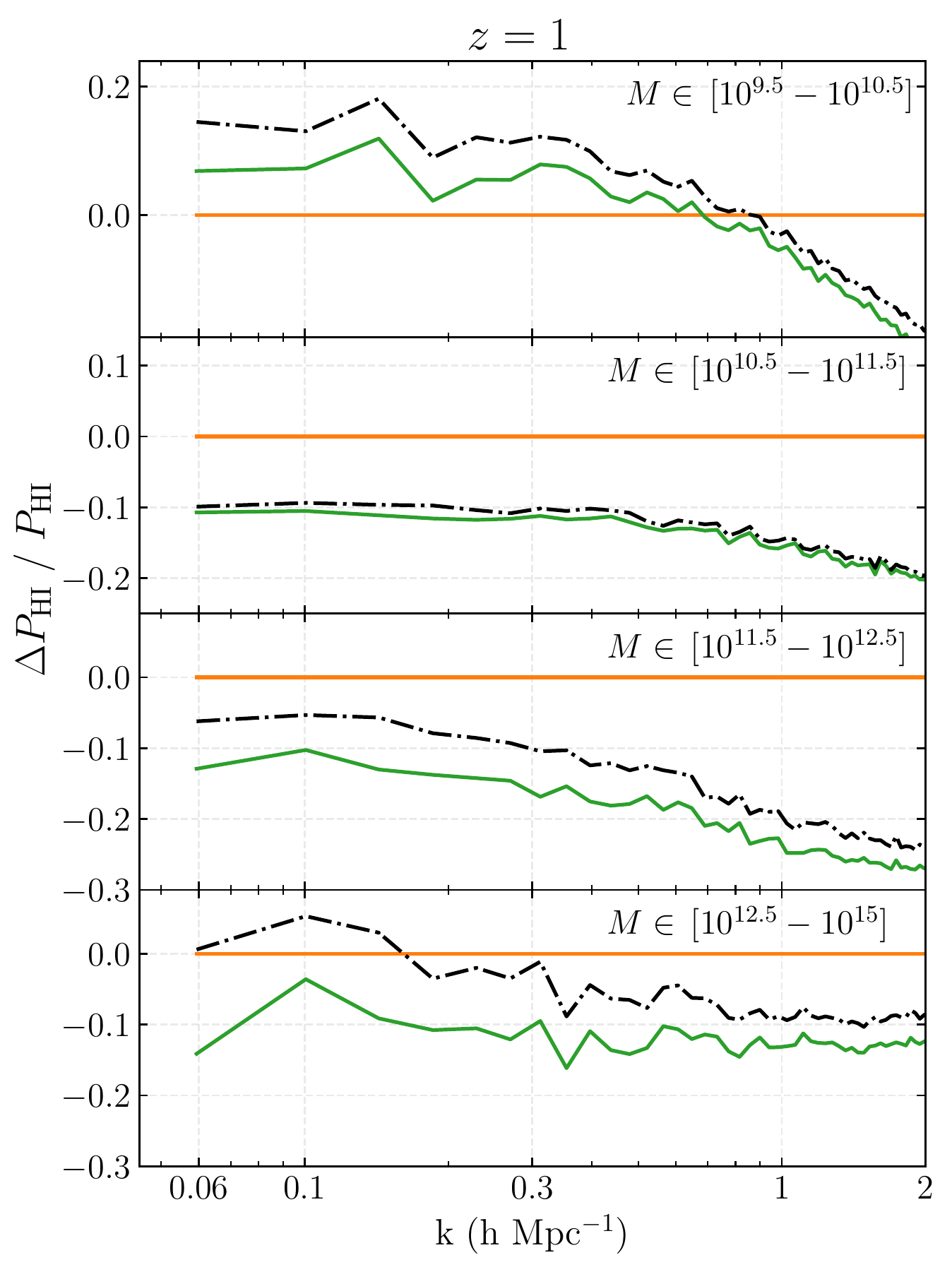}
\caption{Same as the top panel of Fig.~\ref{fig:Pow_AllEnv} but training the random forest (RF) regressors using the subhalo mass fraction of the halos ($M_\textup{subhalos}/M_h$). 
In most cases, we find improvements similar to those seen for environmental parameters in Fig.~\ref{fig:Pow_AllEnv}. Note however that the predictions become worse for the case of low-mass halos in $z=1$; 
see the text for an explanation. We found that training the RF using the halo concentration had a small effect on the $P_\textup{HI}$ predictions, and hence we do not show results for that case.}
\label{fig:Pow_Others}
\vspace{0.5cm}
\centering
\includegraphics[scale=0.5,keepaspectratio=true]{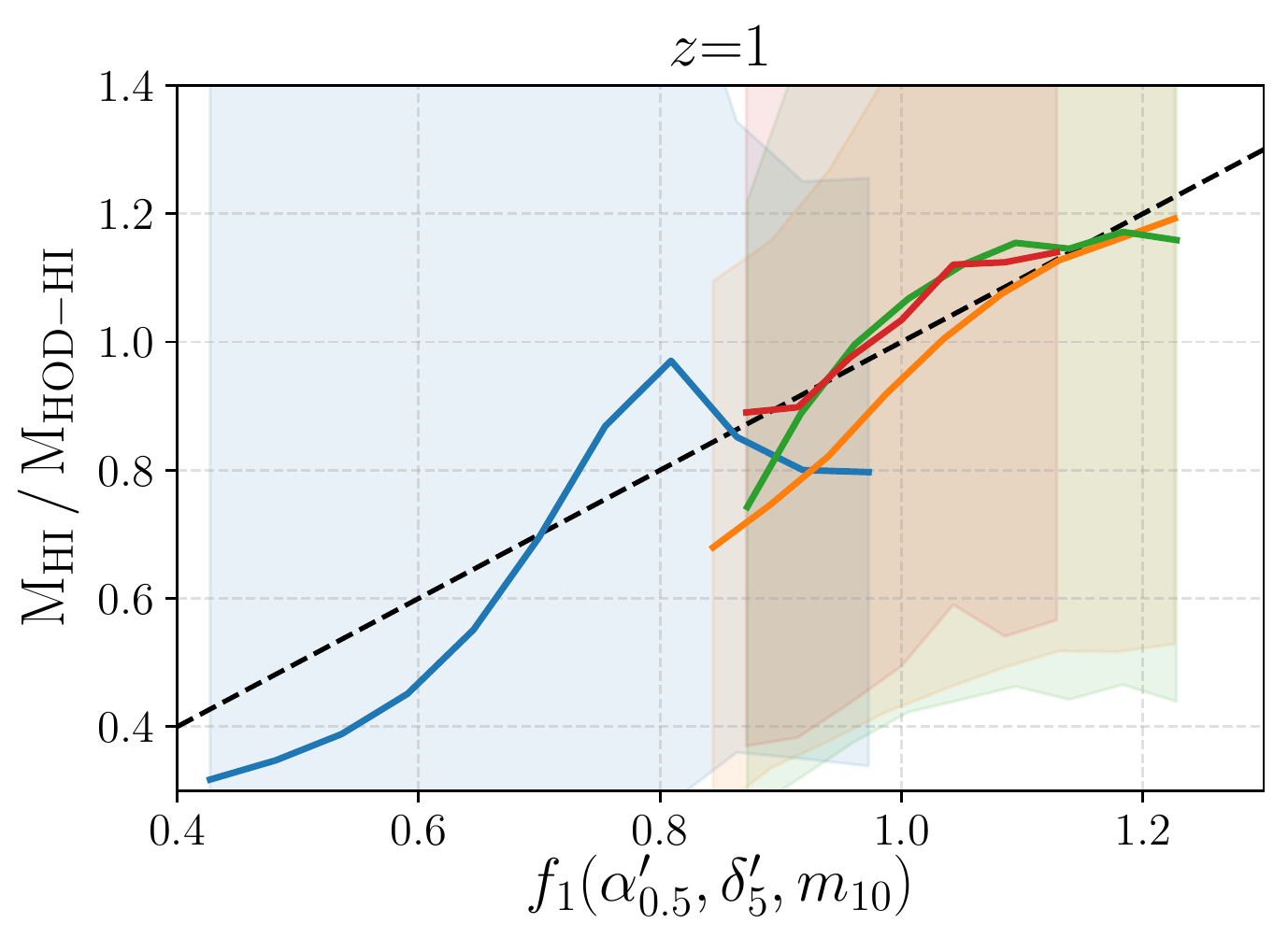}
\includegraphics[scale=0.5,keepaspectratio=true]{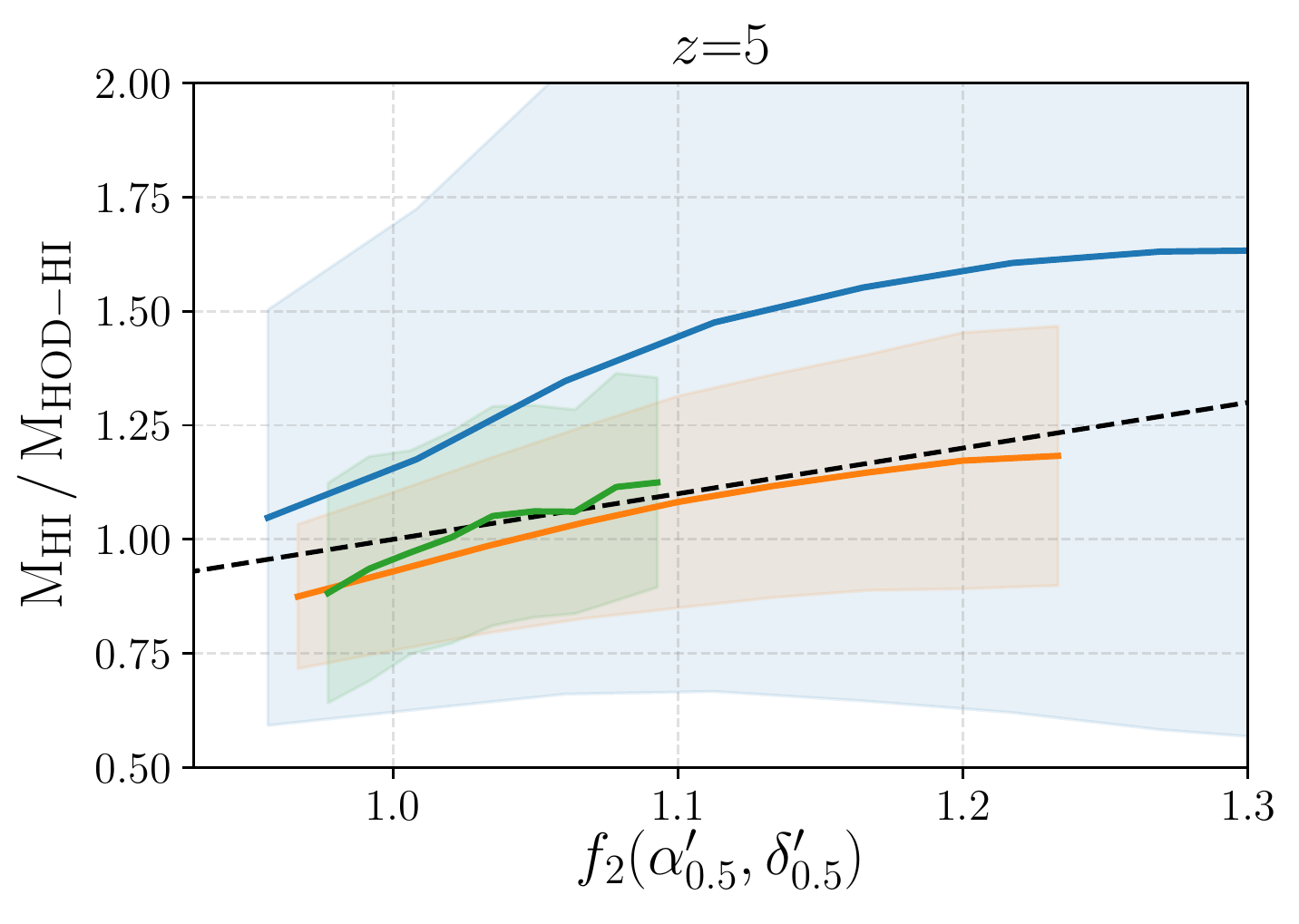}
\caption{We train a symbolic regressor to predict the ratio of the \HI mass of a halo to the output $M_\textup{HOD-HI}$ from the mass-based HOD model in Eq.~\ref{eq:HOD}. We compare the values predicted from Eq.~\ref{eq:SyReg} against the true values from IllustrisTNG (the black dashed line represents predicted=true). $f_1$ and $f_2$ are the expressions in the RHS of Eq.~\ref{eq:SyRegz1} and Eq.~\ref{eq:SyRegz5} respectively. The color coding of the curves is the same as the figure below. Note that the offset seen in the blue line in the right panel is because the HOD model in Eq.~\ref{eq:HOD} does not fit the data well in the low $M_h$ regime for $z=5$. Note that the ratio $M_\textup{HI}/M_\textup{HOD-HI}$ corresponds to the level of assembly bias seen in the hydrodynamic simulation and the symbolic regressor is thus efficient at finding parameter combinations which can be used for model assembly bias.}
\label{fig:mHI_SyReg}
\vspace{0.5cm}
\centering
\includegraphics[scale=0.5,keepaspectratio=true]{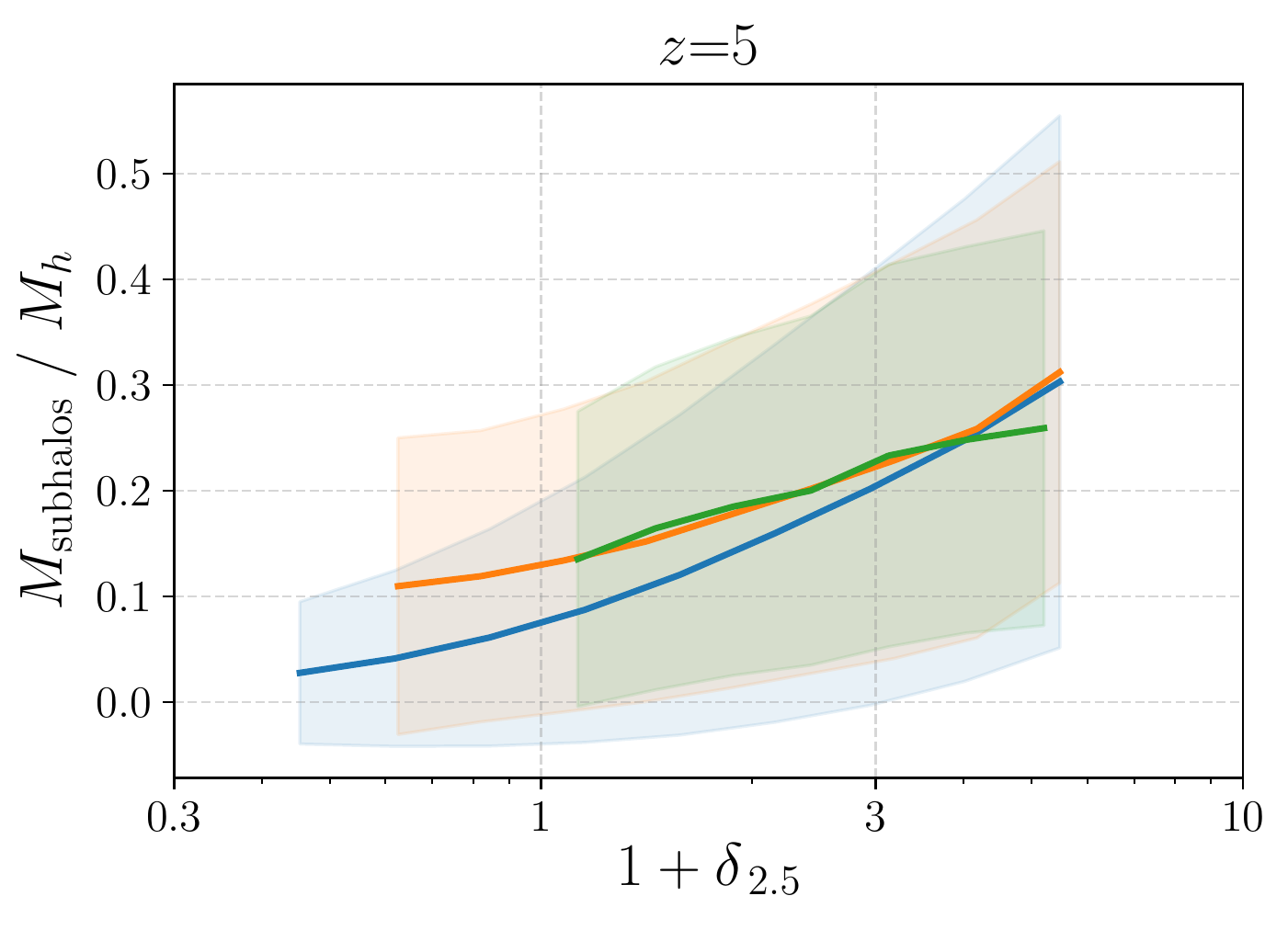}
\includegraphics[scale=0.5,keepaspectratio=true]{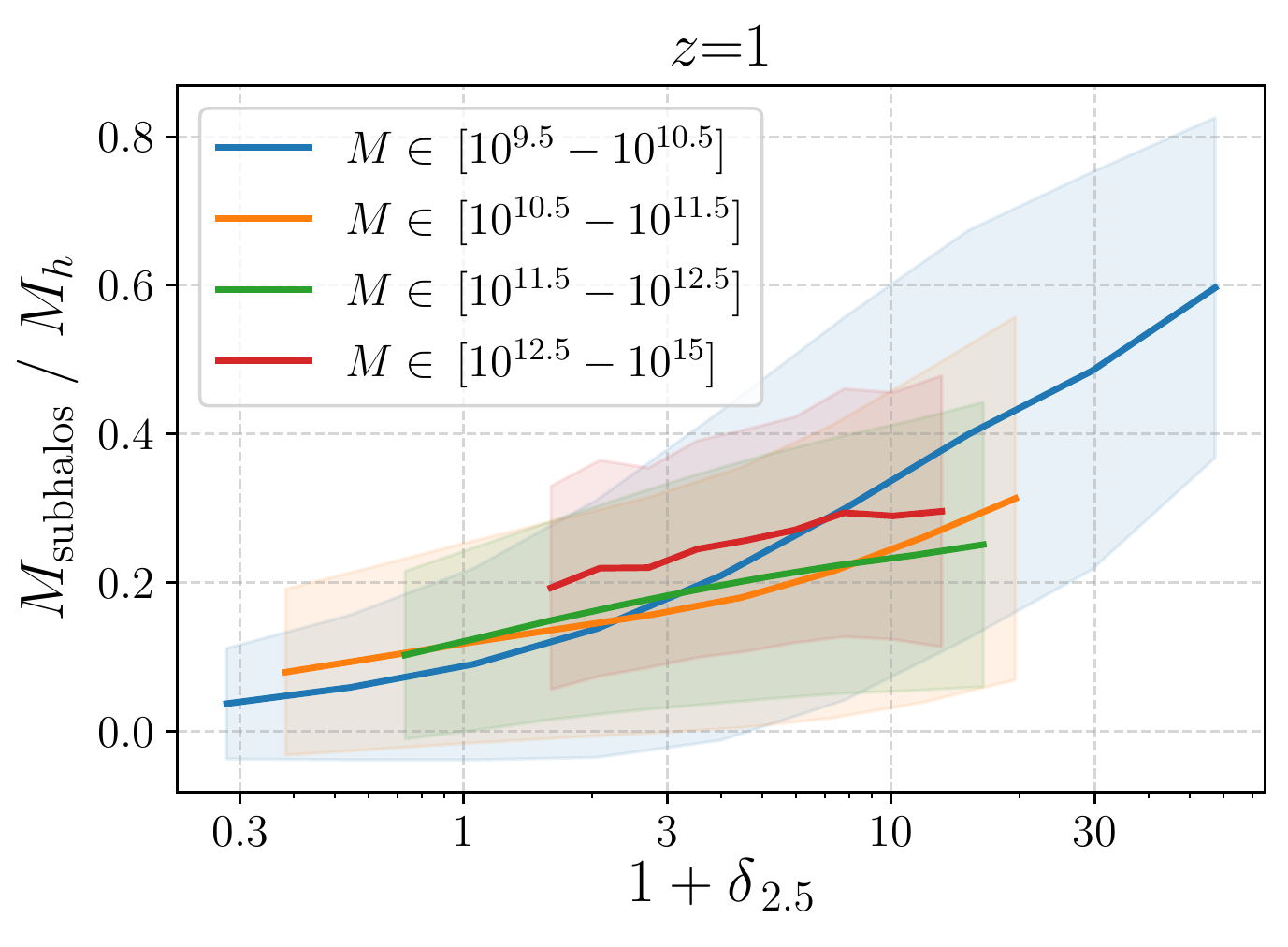}
\caption{Effect of the environmental overdensity of the halo, parameterized by $\delta_{2.5}$, on its subhalo mass fraction. Increasing the environmental overdensity typically results in more mergers and therefore more substructure in halos. These trends are shown to complement the physical explanations outlined in Sec.~3\ref{sec:physical} for the effect of the environment seen on the halo \HI mass in Fig.~\ref{fig:mHI}.}
\label{fig:mHI_others}
\end{figure*}

 We show the ratio of masses of the matched halos in the left panel of Fig.~\ref{fig:DMOvsFP}. We see that the halo masses are different by $\gtrsim$10\% in the two simulations at both high and low redshifts. This difference also affects the HOD modeling of the \HI field. Our model was calibrated for the halo masses from the FP simulation ($M^\textup{FP}_h$). Using this model on halos from the DMO version of the TNG300-1 box, we find that $P_\textup{HI}$ of the modeled field is discrepant, as seen in the purple curves in the center and right panels of Fig.~\ref{fig:DMOvsFP}. We also find that $\Omega_\textup{HI}$ is overpredicted by $\sim 15$ \%. This discrepancy was also encountered by \citetalias{VilGenCas1810} who did a similar analysis for the TNG100-1 box (see their Fig.~24). 
 
In order to appropriately recalibrate the HOD model such that it can be used on a $N$-body simulation, we refit the formula in Eq.~\ref{eq:HOD} using $M_\textup{HI}$ from the FP simulation and $M^\textup{DMO}_h$ from the DMO simulation. We obtain the model parameters $\{\alpha,M_0/(10^{10}\Ms),M_\textup{min}/(10^{10}\Ms)\}$ to be \{0.531, 0.589, 26.285\} for $z=1$ and \{0.727, 0.395, 5.216\} for $z=5$. Using the recalibrated HOD model on the halos from the DMO simulation, we find that the results for $P_\textup{HI}$ are now improved, as seen from the red curves in Fig.~\ref{fig:DMOvsFP}.





\section{Machine learning techniques used}
\label{apx:Techniques}

In the introduction section of the main text, we outlined the motivation of using machine learning models like random forests or symbolic regression instead of deep neural networks. In this appendix we provide more details on these models.

\subsubsection*{Symbolic regression}
Symbolic regression (SR) is a tool that searches the space of mathematical expressions to find the best equation that fits the data. 
The difference between it and ordinary ``least squares'' regression is that knowledge of the underlying functional form of the fitting function is not required a priori. Let us briefly describe the procedure to fit a function (e.g. Eq.~\ref{eq:intro}) with the \textsc{PySR} package.
First we specify the relevant input parameters (e.g., $\delta_{0.5},\alpha_{0.5}$ or $\delta_{5}$) and the operations (e.g., sum ($+$), multiplication($\cdot$), exponential or sinus). Using genetic programming \cite{GenProgram}, the SR then generates multiple iterations of formulae, like $2.7\cdot\delta_{0.5}+\alpha_{0.5}\cdot \delta_{5}$ for example, and outputs a final list of equations which have the lowest mean squared error when compared to the data.
The equations in the final list are ranked on the basis of their complexity (more complex operations like exponentials are penalized over standard ones like $+$). 

Because it is easy to overfit given the large scatter in the simulation (see Fig.~\ref{fig:mHI}) data, we restrict ourselves to the simplest case of sum and multiplication operators. Furthermore, in the final list of formulae obtained from \textsc{PySR}, we choose the most simplest ones to present in Eqs.~\ref{eq:SyReg}. We also show the performance of the expressions in Fig.~\ref{fig:mHI_SyReg}. Note also that dimensionality of the input dataset needs to be relatively small in order for the use of SR to be feasible as the cost scales exponentially with the number of input parameters and operators. We therefore needed to reduce assembly bias dependence to a few most important parameters before using SR.

\subsubsection*{Random forests}
\label{sec:RF}
We use the random forest algorithm from the publicly available package \texttt{Scikit-Learn} \citep{Scikit}.
A random forest regressor (RF) is a collection of decision trees; each tree is in itself a regression model and is trained on a different random subset of the training data \citep{Bre01_RF}. The output from a RF is the mean of the predictions from the individual trees. Note that a single decision tree is prone to overfitting and using the ensemble mean of the different trees helps to reduce overfitting.
RFs have been used for applications to cosmological problems \cite{LucPei18, MosNaa20, NadMao18}, and allow for an easy measurement of the relative importance of each feature. This makes them better suited for interpretation as compared to deep neural networks. 

We train the RFs to model the ratio of the \HI mass of the halo to the prediction from the mass-only HOD model from Eq.~\ref{eq:HOD} ($M_\textup{HI}/M_\textup{HOD-HI}$), as a function of different secondary properties of the halo. We use 100 decision trees in our model. For the hyperparameters which control the complexity of the model (and avoid overfitting), we use: \texttt{max\_depth}=100 and \texttt{min\_samples\_leaf}=10. 

\clearpage



%
%
%
\clearpage
\section*{References}
\bibliography{HI}
\end{document}